\documentclass[fleqn,usenatbib]{mnras}

\usepackage{newtxtext,newtxmath}
\usepackage[T1]{fontenc}

\DeclareRobustCommand{\VAN}[3]{#2}
\let\VANthebibliography\thebibliography
\def\thebibliography{\DeclareRobustCommand{\VAN}[3]{##3}\VANthebibliography}

\usepackage{graphicx}	% Including figure files
\usepackage{amsmath}	% Advanced maths commands
\newcommand{\vlos}{v_{\mathrm{los}}}

\newcommand{\kms}{ \ {\rm{km \ s^{-1}}}\>}

\newcommand{\foo}[1]{}

\newcommand{\hyperfootnote}[1][]{\def\ArgI\hyperfootnoteRelay}
\newcommand\hyperfootnoteRelay[2][]{\href{#1#2}{\ArgI}\footnote{\href{#1#2}{#2}}}
\title[Internal kinematics in TNG50]{Internal kinematics of 
dwarf satellites of MW/M31-like galaxies in TNG50}

\author[A. M. Martínez-García et al.]{
Alberto Manuel Martínez-García,$^{1,2}$\thanks{E-mail: ammtnez@iac.es},
Andrés del Pino,$^{3}$
Ewa L. Łokas, $^{4}$
Roeland P. van der Marel,$^{5,6}$
\newauthor
and Antonio Aparicio $^{1,2}$
\\
$^{1}$Instituto de Astrofísica de Canarias, Calle Vía Láctea S/N, E-38205 La Laguna, Tenerife, Spain\\
$^{2}$Universidad de La Laguna, Dpto. Astrofísica, Avda. Astrofísico Fco. Sánchez S/N, E-38206 La Laguna, Tenerife, Spain\\
$^{3}$Centro de Estudios de F\'isica del Cosmos de Arag\'on (CEFCA), Unidad Asociada al CSIC, Plaza San Juan 1, 44001, Teruel, Spain\\
$^{4}$Nicolaus Copernicus Astronomical Center, Polish Academy of Sciences, Bartycka 18, 00-716 Warsaw, Poland\\
$^{5}$Space Telescope Science Institute, 3700 San Martin Drive, Baltimore, MD 21218, USA \\
$^{6}$Center for Astrophysical Sciences, The William H. Miller III Department of Physics \& Astronomy, Johns Hopkins University, Baltimore, MD 21218, USA\\
}

\date{Accepted 2023 September 21. Received 2023 September 20; in original form 2023 July 25}

\pubyear{2023}

\begin{document}
\label{firstpage}
\pagerange{\pageref{firstpage}--\pageref{lastpage}}
\maketitle

\begin{abstract}

We present a kinematic study of a thousand of dwarf satellites 
of MW/M31-like hosts from the IllustrisTNG50 simulation. Internal 
kinematics were derived for all the snapshots to obtain a 
historical record of their rotation velocity in the plane of the 
sky ($|V_T|$) and the amplitude of their velocity gradients 
along the line of sight ($A_{\rm grad}^{v_z}$) measured from the 
host. For  the majority of the satellites, we initially detected 
rotation in the plane of the sky (65\%) or velocity gradients 
(80\%), and  this was progressively reduced to  45\% and 
68\% at $z = 0$, respectively.  We find that the evolution of the 
rotation in the plane of the sky and the velocity  gradients 
differs according to type of dwarfs, which could be explained in terms of  
their different masses and orbital histories. We observe that 
interaction with the host has an impact on the evolution of 
the  internal kinematics of the satellites. The rotation signal 
of the satellites is progressively reduced during pericentric 
passages, the first pericentre being especially disruptive 
for the initial kinematics. We observe temporary increases in 
$A_{\rm grad}^{v_z}$ during pericentric passage caused by tidal interaction 
with the host, $A_{\rm grad}^{v_z}$ increasing as the satellites 
approach their pericentre and dropping as they move away. In 
summary, we conclude that the presence of detectable rotation in 
dwarf satellites is not uncommon, and that the evolution of their 
internal kinematics is clearly affected by their interaction with 
the host.

\end{abstract}

\begin{keywords}
galaxies: dwarf -- galaxies: evolution  -- galaxies: kinematics and dynamics -- Local Group
\end{keywords}

%%%%%%%%%%%%%%%%%%%%%%%%%%%%%%%%%%%%%%%%%%%%%%%%%%

%%%%%%%%%%%%%%%%% BODY OF PAPER %%%%%%%%%%%%%%%%%%

\section{Introduction}
Dwarf galaxies outnumber any other type of galaxy in the Universe. 
According to the $\Lambda$ Cold Dark Matter ($\Lambda$CDM) model,
 dwarfs were the first galaxies to be formed and are the basic building blocks 
of larger galaxies, which are formed through mergers and gas 
accretion (\citealt{WhiteRees1978, Blumenthal1984, Dekel1986, NavarroFrenkWhite1995}). 
In this scenario present-day dwarfs would not yet have merged
to form large structures and thus could conceal relevant 
information about the early Universe. 

Local Group (LG) galaxies offer an exceptional opportunity to study 
dwarf galaxies, given their abundance and relative proximity 
(\citealt{McConnachie2012}). 
Analysis of the nearest dwarfs provides detailed information 
about their resolved populations, thus making the LG an ideal laboratory 
for understanding dwarfs.
Among the wide range of properties, stellar populations, and 
morphologies shown by dwarf galaxies, there are two prominent 
sub-categories: dwarf irregular galaxies (dIrrs) and dwarf 
spheroidal galaxies (dSphs). DIrrs are gas-rich galaxies with 
recent star formation,
and show coherent kinematics to some extent 
(\citealt{Mateo1998, McConnachie2012, Kirby2014, Weisz2014, Putman2021}). 
On the other hand, dSphs are essentially devoid of gas, 
with no recent star formation, and are dominated by random 
motions (\citealt{Mateo1998, McConnachie2012, Weisz2014, Wheeler2017, Putman2021}). 
In addition to all these properties shared between dSphs and 
dIrrs, their distribution in the LG is also different, dSphs 
being  located close to the largest galaxies of the group and dIrrs in
 the outskirts (\citealt{Mateo1998}).
This suggests the presence of a transformation mechanism, which 
is thought to have converted gas-rich discy dwarfs (resembling 
present-day dIrrs) into gas-poor spheroids, owing to a combination 
of ram-pressure stripping, tidal stirring, and other environmental 
mechanisms (\citealt{Mayer2010} and references therein). These 
processes would remove the gas, stop the star formation, and 
disrupt the internal kinematics of the dwarf. However, according 
to this scenario, resulting dSphs could retain some residual 
rotation (\citealt{Klimentowski2009, Kazantzidis2011, Lokas2015}).

The internal kinematics of dSphs remains one of their most 
intriguing and least-known aspects. 
The only way of performing kinematic analysis in dSphs is through 
the velocities of their individual stars. 
Initially, the kinematic data for dSphs were limited to catalogues
 of line of sight velocities ($\vlos$). In most cases no clear 
evidence of rotation was found in many dSphs 
(\citealt{Kleyna2002, Wilkinson2004, Munoz2005, Munoz2006, Koch2007a, Koch2007b, Walker2009}),  
albeit for Carina, Fornax, Sculptor, Sextans, and Ursa Minor 
some hints of rotation based on line-of-sight velocity gradients
 were reported 
(\citealt{Battaglia2008b, Battaglia2011, Amorisco2012, Fabrizio2016, Zhu2016, delPino2017, Pace2020}). 
However, the lack of reliable  proper motion (PM) measurements did not allow  clarification 
in most cases of whether these gradients were due to actual rotation 
or to projection on the sky of the PMs along the 
line of sight (\citealt{Feast1961}). 

 Data releases of the \textit{Gaia} 
mission (\citealt{GaiaCollaboration2016}) in recent years have offered a 
unique opportunity to study LG satellites using the PMs  of thousands of their stars. This has meant a 
revolution in our understanding of the LG. The presence of 
coherent motions in several dSphs has been reported using PMs 
from \textit{Gaia}. In particular, rotation in the plane 
of the sky has been detected for Carina, Fornax, Sagittarius, 
and Sculptor, and 
velocity gradients along the line-of-sight (not affected by 
projection effects) have been reported for Carina, Draco, 
Fornax, Sagittarius, and Ursa Minor 
(\citealt{delPino2021, MartinezGarcia2021, MartinezGarcia2022}).
This suggests that the presence of coherent motions in the 
internal kinematics of classical dSph satellites of 
the Milky Way (MW) may not be an uncommon feature. However, the sample 
of galaxies for which analysis of internal kinematics 
is feasible at present remains limited. The study of the 
internal kinematics of a dwarf satellite requires a sufficient 
number  of member stars with solid measurements of the PMs 
and $\vlos$. 
This significantly restricts analysis to a small number 
of MW satellites and does not allow us to draw strong conclusions 
regarding the frequency of these phenomena in the LG or elsewhere.

Simulations are an excellent tool for studying the internal 
kinematics of dSph satellites, as they allow us to complement 
and interpret observational results in ways that would not otherwise be 
possible.
In many cases, tailor-made or zoom-in simulations were used to 
reproduce the evolution of a dSph as it interacts with a MW 
analogue or another dwarf under a wealth of scenarios and 
conditions 
(e.g. \citealt{Klimentowski2009, Kazantzidis2011, Lokas2014, 
Lokas2015, Ebrova2015, CardonaBarrero2021}). 
It is useful to understand how the interaction impacts 
the dwarf, whose parameters are more relevant in the evolution, 
and the final outcome of the kinematics. However, this kind of 
simulation reproduces the evolution for a limited number of 
dwarfs and scenarios. Cosmological simulations offer an 
alternative approach, allowing us  to study the evolution of 
large samples of dwarfs under a variety of complex environments. 
Cosmological simulations have proved to be a remarkable tool in 
understanding the abundance (\citealt{Engler2021}), colours 
(\citealt{Sales2015}), and star formation of satellite galaxies 
(\citealt{Joshi2021, Engler2022}). Nevertheless, the study of 
their internal kinematics remains largely unexplored. 
Cosmological simulations can be an effective tool for analysing 
the internal kinematics of dwarf satellites, provided they 
have enough resolution in the baryonic particles and comprise 
large enough cosmological volumes.

In this paper, we present a kinematic study of dwarf satellites 
from the TNG50 run of the IllustrisTNG
suite of simulations (\citealt{Marinacci2018}). We perform an 
analysis similar to that of \citet{MartinezGarcia2021, MartinezGarcia2022} 
for the MW dSph satellites but using simulated TNG50 data. 
This allows us to study the evolution of the internal kinematics 
of simulated satellites from an observational point of view and 
thus compare the results with previous observational studies.
We assess the presence of coherent motions in the internal kinematics 
of dwarf satellites at $z = 0$, and study the evolution of their 
kinematics across time and the role of the host galaxies in the 
evolution of the kinematics.

This paper is organized as follows. In Section~\ref{sec:data}, we 
present data and methods, in Section~\ref{sec:results} we 
present the results, and finally in Section~\ref{sec:conclusions} 
we summarize and present the main conclusions of this work.

\section{Data and methods}
\label{sec:data}
\subsection{Simulations}
To study the internal kinematics of dwarf satellites we use the 
TNG50 run of the IllustrisTNG project 
(\citealt{Marinacci2018, Naiman2018, Nelson2018, Springel2018, Pillepich2018b}). 
The IllustrisTNG project consists of a suite of 
gravo-magnetohydrodynamic simulations that follow the evolution
 of dark matter, stellar particles, black holes, stellar winds, 
and gas cells between redshift $z=127$ and $z=0$. For each run 
there are 100 snapshots available, ranging from from $z \sim 20$ to $z=0$. 
The simulations were run with the moving-mesh code \texttt{AREPO} 
(\citealt{Springel2010}) and adopt cosmological parameters 
consistent with \citet{PlanckCollaboration2016} 
($\Omega_{\Lambda,0} = 0.6911$, $\Omega_{m,0} = 0.3089$, 
$\Omega_{b,0} = 0.0486$, $\sigma_8 = 0.8159$, $n_s = 0.9667$, 
and $h = 0.6774$). As for the model for the formation and evolution 
of galaxies, IllustrisTNG uses an improved version of the Illustris 
model (full details of which can be found in \citealt{Weinberger2017, Pillepich2018a}).

The detection of galaxies is performed following a two-step 
process. First, the haloes are identified with the friends-of-friends 
(FoF; \citealt{Davis1985}) algorithm with linking length $b=0.2$. 
Finally, subhaloes (galaxies) are detected using \texttt{SUBFIND} 
(\citealt{Springel2001, Dolag2009}).  In order to trace the identified subhaloes between the snapshots, the merger trees are built using the \texttt{SUBLINK} algorithm (full description in \citealt{RodriguezGomez2015}), which follows the baryonic content of the galaxies throughout the simulation.

For our study, we selected the TNG50-1 run (hereafter TNG50; 
\citealt{Nelson2019b, Pillepich2019}) of the IllustrisTNG suite. 
TNG50 is the run with the highest resolution in baryonic 
particles ($8.5 \times 10^4 M_{\odot}$) and comprises a significant 
cosmological volume ($\sim 50$ Mpc comoving box). These two 
features ensure that we can find a sizeable number of dwarf 
satellite galaxies with enough stellar particles to perform the 
analysis of their internal kinematics.

\subsection{Selection of dwarf satellites}
\label{subsec:sel}
We aim to select dwarf satellites from TNG50 similar to those 
of the LG. In the first place, we choose host galaxies whose 
halo virial masses are comparable to the virial masses of the 
MW/M31 (0.7--$3.0\times 10^{12} M_{\odot})$ at $z = 0$ (see 
\citealt{Patel2017a} and references therein). We then analyse 
their satellites, having selected those that meet a series of 
criteria, namely: 
\begin{itemize}
    \item The satellite has to belong to the halo of the host galaxy.
    \item It has to be located within the virial radius of the host at $z=0$.
    \item Its maximum circular velocity has to be $\leq 45 \kms$ (\citealt{Patel2018}).
    \item Its stellar mass has to be  $\leq 10^{9} M_{\odot}$ (\citealt{Bullock2017}).
    \item It has to contain at least 50 stellar particles at $z = 0$.
    \item The value of the flag 'SubhaloFlag' (\citealt{Nelson2019a}) 
has to be equal to one. This allows us to discard subhaloes detected 
by \texttt{SUBFIND} that are clumps of baryonic matter rather 
than galaxies of actual cosmological origin. 
    \item The galaxy can be traced back up to at least $z=2$ (\citealt{Joshi2021}).
\end{itemize}

These criteria allow us to obtain a large sample of dwarf satellites 
($\sim10^3$ galaxies) with enough stellar particles to study their 
internal kinematics. In order to analyse the internal kinematics 
of each class of dwarf, we aggregated them as dSphs and non-dSphs 
according to their observed properties in the LG 
(\citealt{Weisz2014, Putman2021}). The classification is based 
on the gas content and star formation histories of the galaxies 
(\citealt{Bullock2017}). We label `dSph' for those dwarfs whose gas 
mass represents less than 1 per cent of their total mass at $z = 0$, 
and that completed 90 per cent of their star formation at $z \geq 0.5$.
 The remaining dwarfs are labelled `non-dSphs' and can be a mixture 
of dIrrs, transition dwarfs, or dwarf ellipticals.

\subsection{Derivation of internal kinematics}
\label{sec:internal_kinematics}
We derive the internal kinematics of the selected dwarf 
satellites for every snapshot of TNG50. 
For every halo and its corresponding satellites we use 
a galactocentric system whose origin is located at the 
centre of the host. We define the centre of a galaxy as 
the position of its most bound particle. From the centre 
of the host we observe the kinematics of the satellites. 
For  all the stellar particles assigned to each subhalo by 
the \texttt{SUBFIND} algorithm, we derive their coordinates, 
PMs, and distances measured from the centre of the host.
We then derive the internal kinematics of each of the 
satellites following a procedure analogous to that used 
by \citet{delPino2021, MartinezGarcia2021} and 
\citet{MartinezGarcia2022}. We introduce a co-moving 
frame centred on each satellite. The frame is fully described 
by \citet{VanderMarel2001} and \citet{VanderMarel2002}. We 
refer the reader to \citet{delPino2021, MartinezGarcia2021}, 
and \citet{MartinezGarcia2022} for further details on the 
reference frame and its application to dSph satellites of 
the MW. In a nutshell, the 3-dimensional motion of the centre 
of the satellite galaxy is subtracted from its velocity field. 
This bulk motion of the centre is derived by taking the median 
of the velocity components of the stellar particles enclosed 
within the stellar half-mass radius of the galaxy. The 
individual 3D relative velocities of the stellar particles 
are decomposed into three orthogonal components:

\begin{equation}\label{eq:v1v2v3}
v_{S,i} \equiv \frac{dD_i}{dt},\\ v_{R,i} \equiv D_i\frac{d\rho_i}{dt},\\ v_{T,i} 
\equiv D_i\sin{\rho_i}\frac{d\phi_i}{dt},
\end{equation}  
where $D_i$ is the distance of the $i$-th stellar particle 
to the observer (the centre of the host), $\rho_i$ is the angular 
distance to the centre of the satellite, and $\phi_i$ is the position 
angle (measured north to east). 
Consequently, $v_{S,i}$ is the line-of-sight component of the 
velocity, and $v_{R,i}$ and $v_{T,i}$ are the radial and tangential 
components of the velocity in the plane of the sky with respect 
to the centre of the satellite. We derive three additional 
orthogonal velocity components referred to the centre of the 
satellite: $v_{x,i}$, $v_{y,i}$, $v_{z,i}$, where $x$-axis is 
antiparallel to the RA axis; $y$-axis is parallel to the Dec axis; 
and $z$-axis points towards the observer for a stellar particle 
located at the centre of the satellite. Figures 1 and 2 of 
\citet{VanderMarel2002}, and figure 2 of \citet{MartinezGarcia2022} 
provide further clarification on the velocity components and 
reference frame.

This procedure is applied to all the selected satellites in 
each of the snapshots, so we obtain a historic record of 
their internal kinematics. However, in our analysis we only 
consider the internal kinematics of a satellite at a given 
snapshot if it meets certain criteria. The satellite has to 
contain at least 50 stellar particles in order to have a 
sufficient amount of data to study the internal kinematics 
and obtain statistically sound results. Additionally, the median 
value of $v_{S,i}$ (henceforth $V_S$) of its stellar particles 
has to be compatible with zero at $1\sigma$ level in that 
particular snapshot. This implies that an optimal subtraction 
of the bulk motion of the centre of the satellite has been 
performed,  and that its internal kinematics can therefore be trusted. 
We define the error of ${V_S}$ as the standard error of the median 
of $v_{S,i}$, $1.253 \,\sigma_{S} / \sqrt{N_*}$, 
where $\sigma_S$ is the standard deviation of $v_{S,i}$ and $N_*$ is the 
number of stellar particles. In order to verify the validity 
of this error for the median, we performed multiple tests 
using bootstrap simulations to derive $V_S$. We find that the 
standard deviation of the median of the realizations of the 
simulations is similar to the assumed error of $V_S$. 
For a satellite in a certain snapshot meeting  these requirements 
we give full consideration to the measurement of its rotation 
velocity in the plane of the sky and the presence of gradients 
in $v_z$ in that particular snapshot.

We stress that in this study we analyse the internal kinematics 
of the simulated satellites from an observational point of view, 
i.e.\ as would be witnessed by an observer  placed at the centre 
of the host galaxy. The internal kinematics of a 
satellite are therefore going to be observed as a combination of motions 
in the plane of the sky and along the line of sight.
We derive the rotation velocity in the plane of the sky of a 
satellite (henceforth $V_T$) by taking the median of the tangential 
component of the velocity ($v_{T,i}$) of the individual stellar 
particles. We consider a galaxy to be rotating in the plane of 
the sky if $V_T$ is not compatible with zero at the $1\sigma$ level. 
 The error in $V_T$ is defined analogously to that of $V_S$, 
and the same tests with bootstrap simulations have been performed 
to verify the validity of the assumed error for $V_T$. Net positive 
values of $V_T$ mean that the galaxy rotates in an anticlockwise 
direction in the plane of the sky. 

For the detection of velocity gradients in the satellites, 
we fit a plane to the positions $(x,y)$ and the velocity 
component $v_z$  of the stellar particles of each dwarf and 
snapshot using ordinary least-squares regression. The plane is
 defined as $v_z = a_1\,x + a_2\,y + b$. The amplitude of the 
gradient ($A_{\rm grad}^{v_z}$) is the square root of the quadratic 
sum of the fitting coefficients, $\sqrt{a_1^2 + a_2^2}$.
We consider a galaxy to show a gradient in  $v_z$, when 
$A_{\rm grad}^{v_z}$ is not compatible with zero at $1 \sigma$ level.

\begin{figure}
    \centering
    \includegraphics[width=\linewidth]{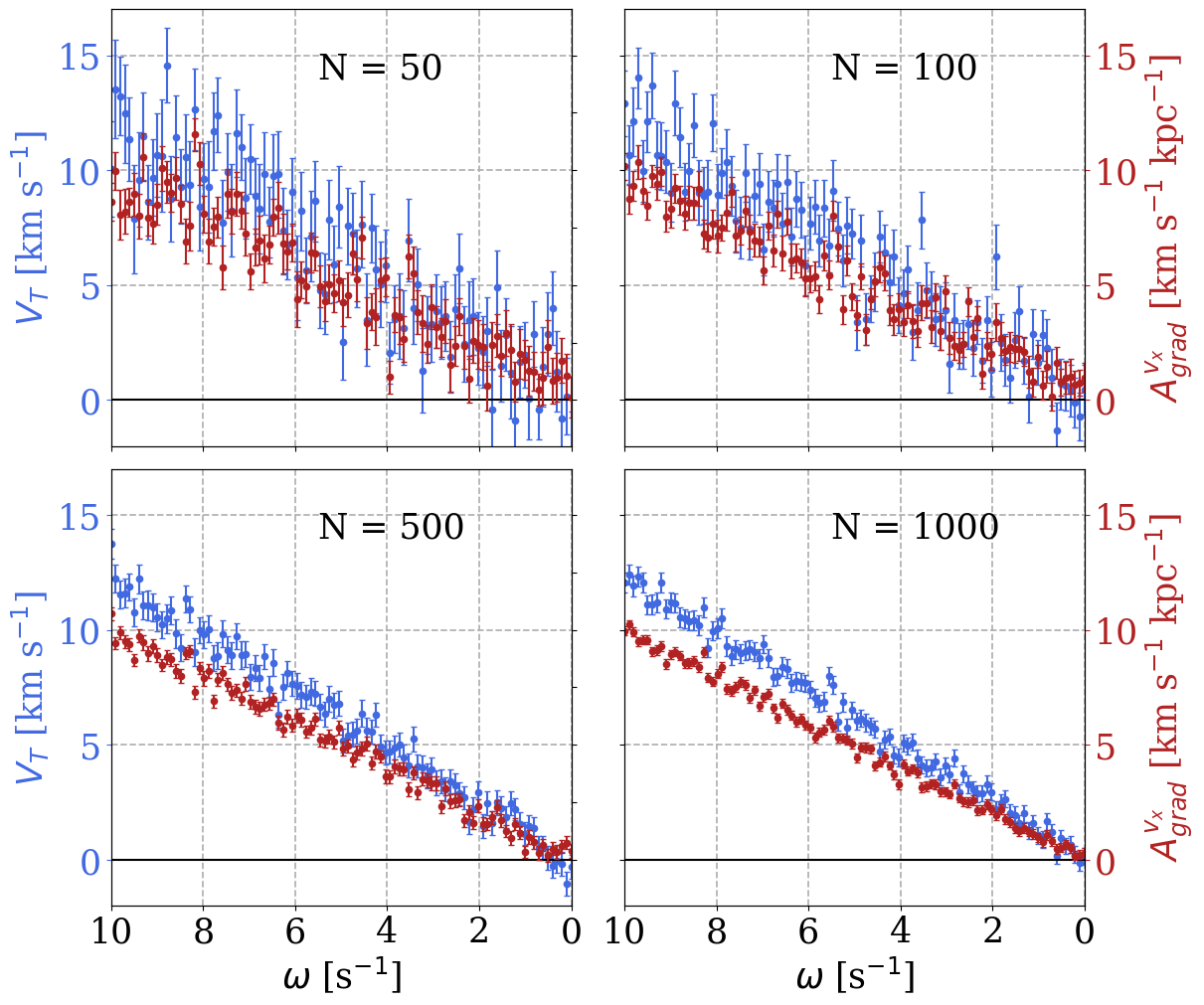}
    \caption{Sensitivity to the detection of rotation as 
rotation in the plane of the sky and as velocity gradients 
for galaxies with different angular velocities and numbers of 
stellar particles. Blue points with error bars represent the 
recovered rotation velocity in the plane of the sky for the 
mock galaxy rotating at a given angular velocity, and red points 
with error bars represent the recovered amplitude of the velocity 
gradient in $v_x$. The panels represent mock galaxies of 50, 
100, 500, and 1000 stars. Note that we detect rotation in the 
plane of the sky when $|V_T| \neq 0$ at $1\sigma$ level and 
gradients in $v_z$ when $A_{\rm grad}^{v_z} \neq 0$ at $1\sigma$.}
    \label{fig:sensitivity}
\end{figure}

We performed multiple tests to assess the capability of 
detecting rotation in the plane of the sky and gradients 
in $v_z$ in galaxies with different numbers of stellar 
particles and rotation velocities. We generated a mock spherical 
galaxy for which the position of the stellar particles was 
sampled from a random normal distribution centred on zero and a 
dispersion equal to 1 kpc. This produces a galaxy whose size 
is comparable to that of a dSph such as Ursa Minor 
(\citealt{Irwin1995}). The galaxy rotates as a solid body 
around the $z$-axis with constant angular velocity ($\omega$), 
the linear velocity of a particle being  sampled from a normal 
random distribution centred on $\omega R$, where $R$ is the 
distance of the particle to the centre of the galaxy, and the 
dispersion is equal to $9.1 \kms$ (the median velocity dispersion 
of the MW classical dSphs, \citealt{McConnachie2012}). We 
measure the rotation velocity in the plane of the sky. 
In order to measure the amplitude of the gradient generated 
by the same rotation signal we then measure the gradient in $v_x$ 
from the plane $y-z$ ($A_{\rm grad}^{v_x}$). This procedure is 
repeated for 100 angular velocities between 0 and 10 s$^{-1}$ 
in mock galaxies of 50, 100, 500, and 1000 stellar particles. 
In Figure~\ref{fig:sensitivity} we show the recovered $|V_T|$
 and $A_{\rm grad}^{v_x}$ for the different angular velocities and 
numbers of stellar particles. We observe how the uncertainties 
are reduced with the number of stellar particles, given that 
they are inversely proportional to the square root of the 
number of stellar particles. This eases the detection of 
rotation in massive dwarfs. Additionally, we found that  
for galaxies with low angular velocities, it is easier to 
detect rotation 
as gradients rather than as rotation in the plane of the sky. 
This implies that the procedure to detect gradients in $v_z$ 
is more sensitive to low rotation signals. 

\section{Results and discussion}
\label{sec:results}
\subsection{Dwarf satellite sample}
\label{sec:sample}
\begin{figure}
    \centering
    \includegraphics[scale=0.17]{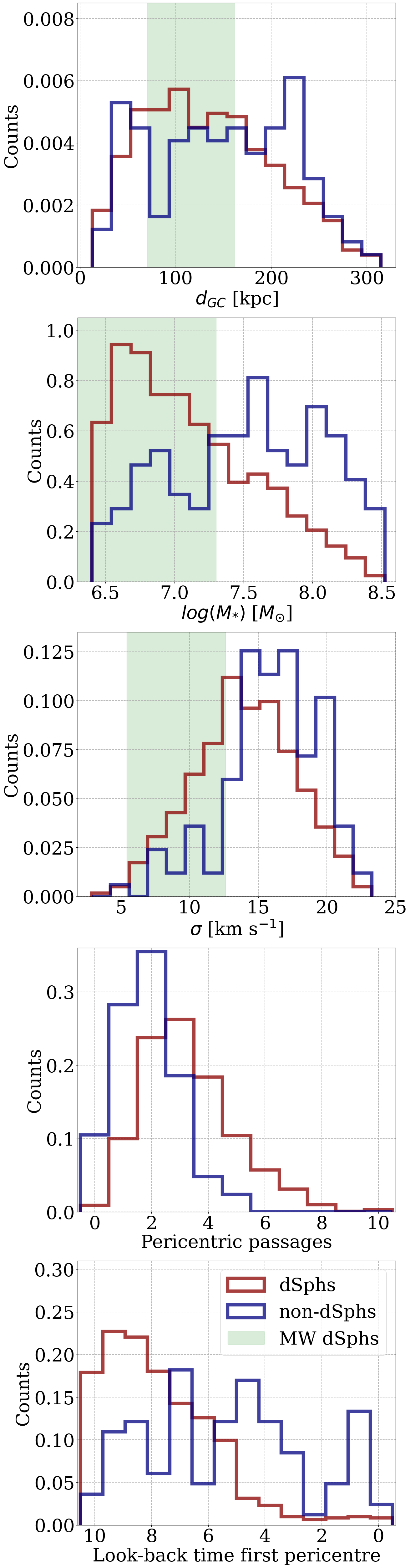}
    \caption{Properties of the selected dwarf satellites at 
$z = 0$. The histograms show, from top to bottom, the 
distribution of the galactocentric distance of the satellites 
to their hosts, the stellar masses, the line-of-sight velocity 
dispersion, the number of pericentric passages, and the look-back 
time of their first pericentric passage for dSphs (marked in 
red) and non-dSphs (marked in blue). The green-shaded areas 
represent the span of the corresponding properties for the 
classical dwarfs of the MW (\citealt{McConnachie2012}).}
    \label{fig:properties}
\end{figure}

We selected 1017 TNG50 dwarf satellites by following the 
criteria listed in Section~\ref{subsec:sel}.  
893 were labelled dSphs and 124 non-dSphs. In 
Figure~\ref{fig:properties}, we show the distribution of 
some of their properties at $z = 0$, namely the distance 
to their host ($d_{\rm GC}$), the stellar mass, the line-of-sight 
velocity dispersion ($\sigma$), the number of pericentric 
passages that they have experienced during the simulation, 
and the look-back time of their first pericentre. 
Both types of dwarfs show similar distributions in $d_{\rm GC}$, 
with non-dSphs having slightly larger values. 
Non-dSph galaxies generally have higher stellar masses 
and higher velocity dispersion. This indicates that the 
mass, both stellar and dynamical,  of non-dSphs tends to 
be larger than that of dSphs.
The analysis of the orbital histories reveals that dSphs 
have endured more pericentric passages (usually between 
1 and 5) than non-dSphs (between 0 and 3), and that their 
first pericentric passage usually took place earlier. 
Therefore, dSphs are likely to have been more affected 
by interaction with the host.

\subsection{Evolution of the internal kinematics}
\label{sec:evointkin}
We followed the evolution of the internal kinematics of 
the selected satellites of TNG50 backwards from $z = 0$. 
We derived $V_T$ and $A_{\rm grad}^{v_z}$ in each snapshot. 
Dwarfs that at a given snapshot show a poor subtraction 
of the motion of its centre or do not contain at least 
50 stellar particles (see Section~\ref{sec:internal_kinematics}) 
are not considered for further study in that particular 
snapshot in order to preserve the quality of the analysis.

\begin{figure}
    \centering
    \includegraphics[width=\linewidth]{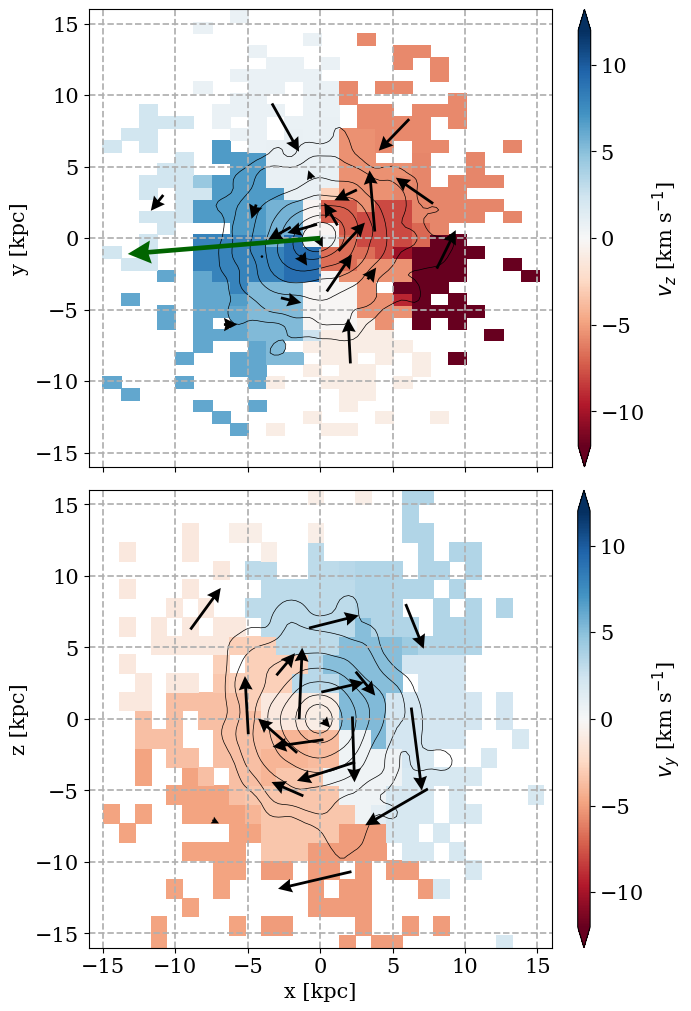}
    \caption{Example of a dwarf satellite (\texttt{subhaloId} 
= 452985) for which we detect rotation in the plane of the sky  
and  a gradient in $v_z$ at $z = 0$. The galaxy contains 5518 
stellar particles. We applied  Voronoi tesellation with cells 
of $\sim$250 particles. The top panel represents the plane $x-y$ 
(i.e.\ the plane of the sky), the bottom panel the plane $y-z$. 
Black arrows represent the median velocity components 
within Voronoi cells. In order to represent the Voronoi cells and their median $v_z$ and $v_x$ in the planes $x-y$ and $y-z$, respectively, we assign the median velocity components of each cell to all the stellar particles belonging to that particular cell. Then we bin the velocities using a rectangular grid. This significantly eases the visualization, giving a continuous appearance to the cells  that is not possible to obtain representing the individual stellar particles. The colour map represents the median $v_z$ 
within cells for the top panel and the median $v_y$ for the bottom 
panel. The green arrow represents the direction of the gradient in 
$v_z$. The projected density of the stellar particles is shown 
by black contours.}
    \label{fig:examples_motions}
\end{figure}

In Fig.~\ref{fig:examples_motions}, we show an example of 
a galaxy for which we detect rotation in the plane of the 
sky and a gradient in $v_z$ at $z = 0$. \texttt{SubhaloId} = 
452985 was chosen as an example given its large number of 
stellar particles (5518), which allows to show its 
kinematic patterns clearly. We show its internal kinematics in the 
$x-y$ plane (i.e.\ in the plane of the sky, top panel) and 
in the $x-z$ plane (bottom panel). We applied Voronoi 
tessellation (\citealt{Capellari2003, delPino2017}) in 
order to increase the signal-to-noise ratio and ease the 
visualization of the coherent motions. The kinematic 
patterns can easily be seen in the corresponding panels. 
In the $x-y$ plane the direction of the rotation is 
consistent with the calculated $V_T$ ($V_T = 2.5 \pm 0.3 \kms$). 
The direction of the gradient in the figure is also consistent 
with the direction of the calculated gradient (PA = 
$94 \pm 4$ deg; measured north to east). Note how the 
gradient in $v_z$ shown in the $x-y$ plane translates into
 a clear clockwise rotation pattern in the $x-z$ plane. 
Velocity gradients in $v_z$ can be caused by the rotation 
of the galaxy along the line of sight. Hence gradients in 
$v_z$ can be considered as an indicator of the presence of 
rotation in perpendicular direction to the plane of the 
sky, as in this case, but could also be caused by tidal forces.

\begin{figure}
    \centering
    \includegraphics[width=\linewidth]{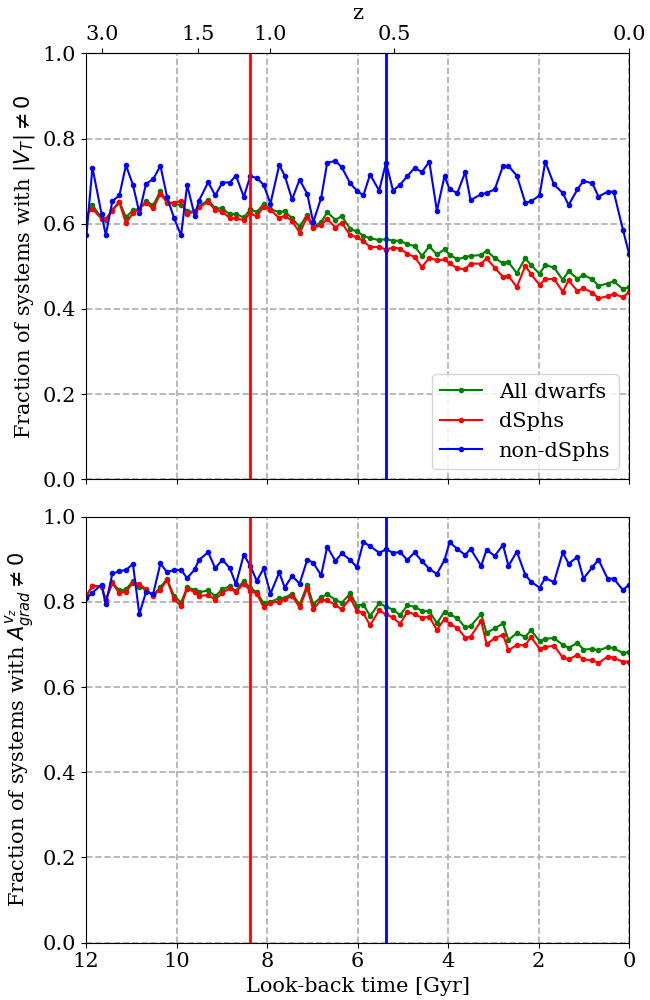}
    \caption{Evolution of the fraction of galaxies for which we detect 
rotation in the plane of sky (top panel) and velocity gradients 
along the line of sight (bottom panel). The green line represents 
the evolution of all the satellites, the red line the evolution of
 dSphs, and blue line the evolution of the remaining galaxies. 
Vertical lines mark the median look-back time for which dSphs 
and non-dSphs performed their first pericentric passages (colours 
match the corresponding type of galaxy).}
    \label{fig:evolution_fraction}
\end{figure}

We detected significant rotation in the plane of the sky 
($|V_T| \neq 0$) and gradients in $v_z$ ($A_{\rm grad}^{v_z} \neq 0$) 
in a sizeable number of systems throughout the simulation. 
The percentage of dwarfs  for which we detect rotation in the 
plane of the sky and gradients in $v_z$ does not remain constant, 
on the contrary, we observe a reduction with time.  
In Fig.~\ref{fig:evolution_fraction}, we show the fraction of 
satellites with detected rotation in the plane of the sky (top 
panel) and the fraction of satellites with detected gradients 
in $v_z$ (bottom panel) at different look-back times/redshifts. 
These ratios are represented for all the galaxies (green line)
 and  are  also represented independently for dSphs (red line) 
and non-dSphs (blue line).

The fraction of galaxies with detected $|V_T| \neq 0$ has 
clearly decreased with time, from $\sim$65\% to $45$\% at 
$z = 0$. 
Initially, dSphs and non-dSphs showed similar ratios. However,
 those ratios started diverging around look-back time $\sim$10 
Gyr. At $z = 0$, we detect rotation in the plane of the sky in  
45\% of dSphs and 53\% of non-dSphs.
The lower mass of dSphs at $z = 0$ (Fig.~\ref{fig:properties})
 makes the detection of rotation in the plane of the sky more 
difficult than for the heavier non-dSphs, which could explain 
the different ratios to some extent. However, we find evidence
 that the difference in the ratios is more related to the different 
evolution of dSphs and non-dSphs than to a detection bias. We 
analysed the ratios of systems with $|V_T| \neq 0$ for dSphs 
and non-dSphs with comparable amounts of stellar particles. 
We note that galaxies with comparable amounts of stellar particles
have also comparable stellar masses, given that the mass of the 
stellar particles is roughly the same.
We divided the satellites using brackets of 250 stellar particles, 
i.e.\ grouping those  containing between 0 and 250 particles, 
between 250 and 500, and so on. We estimated the ratio of dSphs 
and non-dSphs with $|V_T| \neq 0$ within each bracket and found 
that the ratio is generally lower for dSphs than for non-dSphs. 
This suggests that the overall difference in the ratios of dSphs 
and non-dSphs is due to an intrinsic difference in their internal kinematics. 
We applied the Kolmogorov-Smirnov (KS) and Anderson-Darling 
(AD) tests to check the null hypothesis of samples coming from 
the same distribution at $z = 0$. We obtained p-values 0.10 and 
0.04 for the KS and AD tests, respectively. Thus,  the rotation
 velocities in the plane of the sky of dSphs and non-dSphs are 
unlikely to belong to the same distribution. In addition to 
these differences, we note that the fraction of systems with 
$|V_T| \neq 0$ declines earlier and faster for dSphs than for 
non-dSphs, for which the ratio remains roughly constant, with a 
small decrease in the final snapshots of the simulation.
All these differences seem to be connected to the different 
orbital histories and masses of dSphs and non-dSphs.
The earlier decrease in the dSphs could be explained by 
their first pericentric passages taking place earlier (on 
average, $\sim$8.5 Gyr ago) than for non-dSphs ($\sim$5.5 
Gyr ago, considering non-dSphs that do perform at least one passage). 
The fact that dSphs and non-dSphs initially showed similar 
ratios  and have evolved differently depending on their orbital 
histories suggests that the interaction with the host plays a 
role in the evolution of the internal kinematics.
The  different final velocity distributions and the faster 
decrease of the fraction for dSphs are probably due to 
their larger number of pericentric passages and lower mass 
compared to non-dSphs (see Fig.~\ref{fig:properties}). This 
leads to a more intense reduction of the rotation in the plane 
of the sky because  of a more extensive interaction with the 
host in systems that are more sensitive to perturbation, owing 
to their lower mass.

The fraction of satellites with detected gradients in $v_z$ 
has also decreased with time. Initially, $\sim$80\% of satellites 
displayed gradients in $v_z$ and this fraction has been reduced 
down to present-day 68\% (with 66\% for dSphs and 84\% for 
non-dSphs).  
As we did for $V_T$, we compared the ratio of systems with 
gradients in $v_z$ for dSphs and non-dSphs with  similar amounts
 of stellar particles, finding that the ratio is always lower for dSphs.
In the case of the amplitude of the gradients at $z = 0$, we 
applied the KS and AD tests, and  obtained p-values of 0.0005 and 
0.001, respectively, so the distributions of the amplitude of 
the gradients between dSphs and non-dSphs may also be different. 
The fraction evolves differently depending on the category of 
dwarfs,
similarly to what we observe for the fraction of systems rotating 
in the plane of the sky.

Our subsequent analysis  of individual galaxies reveals that the reduction of the ratios is 
related to the evolution of the internal kinematics of the 
satellites (see Section~\ref{sec:evo_v_3}). However, we note that other effects may also have 
some impact on the evolution of the ratios. 
The uncertainties of $V_T$ and $A_{\rm grad}^{v_z}$ are inversely 
proportional to the square root of the number of stellar particles. 
As satellites orbit their host, there is a progressive mass-loss and thus a reduction in the number of stellar particles, 
leading to more difficult detection of $|V_T| \neq 0$ or 
$A_{\rm grad}^{v_z} \neq 0$ at $ 1\sigma$ level with time.  This 
could explain to some extent the reduction in the ratios, yet 
the analysis shows that the actual reduction of the rotation signal 
is the main driver of this behaviour. This will be further studied 
in the following sections.

\subsection{Evolution of the internal kinematics in individual satellites}
\label{sec:evo_v_3}

\begin{table*}
    \centering
    \caption{Properties of the selected satellites of 
Fig.~\ref{fig:evolution_rotation}. Columns: (1) \texttt{subhaloID}, 
(2) total mass, (3) stellar mass, (4) number of pericentric passages, 
(5) distance to the host at $z = 0$, (6) pericentric distance of the 
last pericentre,  (7) look-back time of the last pericentre, (8) 
rotation velocity in the plane of the sky at $z = 0$, (9) amplitude 
of the gradient in $v_z$ at $z = 0$, and (10) the velocity dispersion 
along the line of sight at $z = 0$.}
    \begin{tabular}{c|c|c|c|c|c|c|c|c|c}
    \hline
         \texttt{SubhaloID} & $M$ & $M_*$ & $n_p$ & $d$  & $d_p$ & $t_p$ & $|V_T|$ & $A_{grad}^{v_z}$ & $\sigma$ \\
                            & ($10^8 M_{\odot}$) & ($ 10^6 M_{\odot}$) &  & (kpc) & (kpc) & (Gyr) & ($\kms$) & ($\kms$kpc$^{-1}$) & ($\kms$)\\
         \hline
         \hline
         428192 & 21.3 & 163.9 & 1 & 61 & 28 & 4.6 & 0.62 $\pm$ 0.46 & 2.11 $\pm$ 0.33 & 19.03\\
         
         554807 & 3.8 & 9.2 & 2 & 53 & 24 & 0.1 & 0.06 $\pm$ 1.10 & 2.17 $\pm$ 0.63 & 11.46\\
         516783 & 4.6 & 8.2 & 3 & 194 & 24 & 2.5 & 1.12 $\pm$ 1.18 & 0.57 $\pm$ 0.49 & 10.33\\
         443099 & 1.5 & 55.3 & 6 & 137 &  42 & 0.5 & 0.46 $\pm$ 0.44 & 0.43 $\pm$ 0.50 & 11.72\\
         \hline
    \end{tabular}
    
    \label{tab:prop_examples}
\end{table*}

\begin{figure*}
    \centering
    \includegraphics[width=\linewidth
    ]{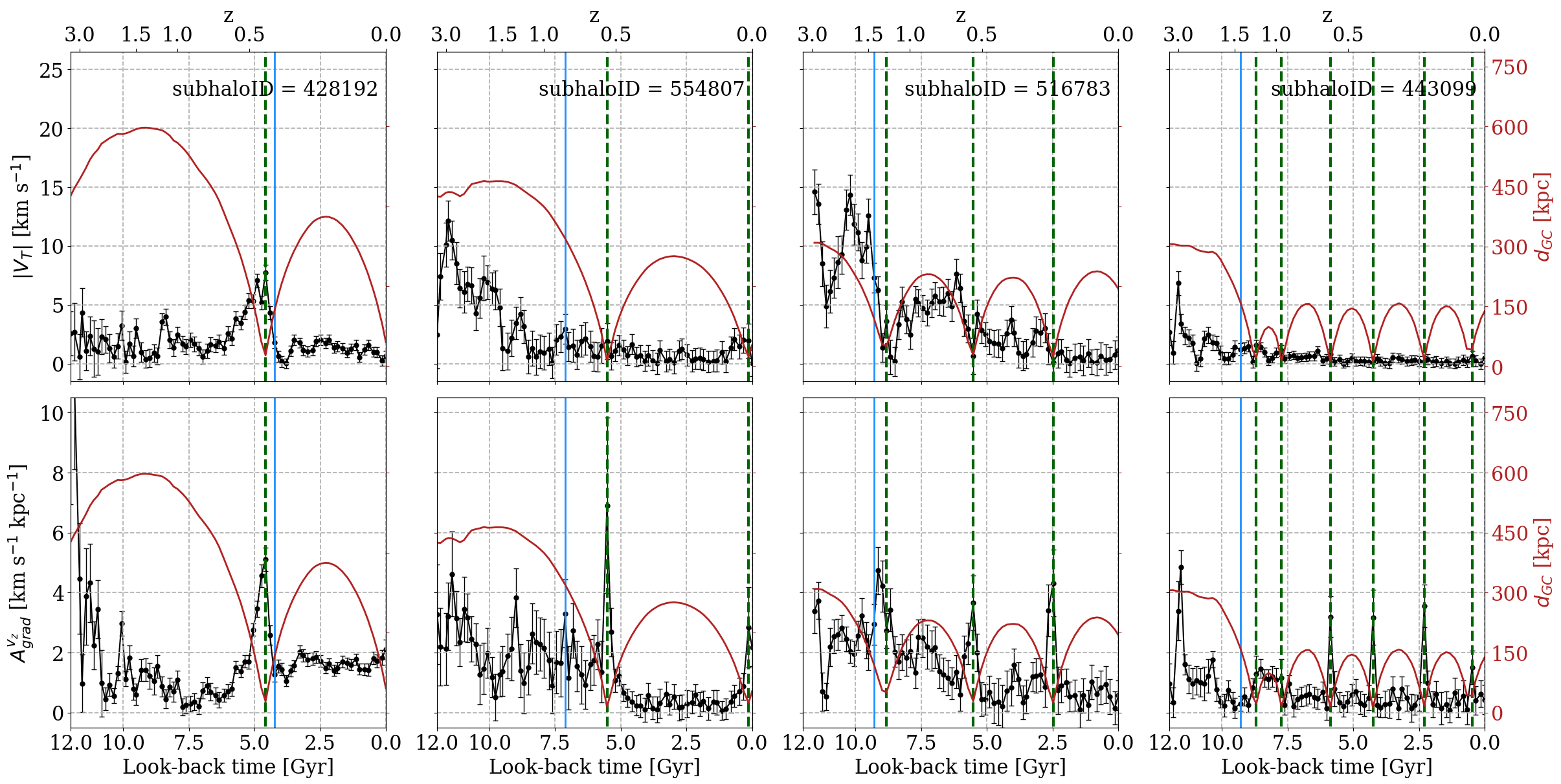}
    \caption{Evolution of $|V_T|$ and $A_{\rm grad}^{v_z}$ for 
four dwarf satellites of TNG50 with pericentric distances 
between $\sim$10 and $\sim$50 kpc. Black lines with error 
bars represent $|V_T|$ (top panels) and $A_{\rm grad}^{v_z}$ 
(bottom panels) at different look-back times/redshifts. Red 
solid lines represent the galactocentric distance of the satellite 
to the host, the pericentric passages are represented by vertical, 
green, dashed lines. Vertical blue lines represent the look-back 
time/redshift at which the satellite reached 90\% of the star 
formation. There is clear reduction of $|V_T|$ and $A_{\rm grad}^{v_z}$ 
associated with the pericentric passages. In the case of 
 $A_{\rm grad}^{v_z}$, we observe repeated increases when the galaxies
 approach the pericentres followed by sharp drops.}
    \label{fig:evolution_rotation}
\end{figure*}

We explored how $|V_T|$ and $A_{\rm grad}^{v_z}$ evolve over 
time in individual dwarfs of our sample in order to have 
a more accurate picture of the impact of the interaction 
with the host on the internal kinematics of the satellites. 
In Fig.~\ref{fig:evolution_rotation}, we show $|V_T|$ and 
$A_{\rm grad}^{v_z}$ at different redshifts for four  satellites 
of TNG50. The selected examples include galaxies with a 
single and multiple pericentric passages and pericentric 
distances between $\sim$10 and $\sim$50 kpc. Their properties 
can be found in Table~\ref{tab:prop_examples}.
We observe that initially, satellites usually have larger 
rotation velocities in the plane of the sky and gradients 
in $v_z$ with larger amplitudes that tend to be reduced with time. 
As for the evolution of $|V_T|$, we generally observe a decrease
 in the average rotation velocity associated with pericentric
 passages. In general, the first pericentre seems to play a main
 role in the transformation of the rotation patterns in the plane 
of the sky. Tidal interaction with the host deeply distorts
 the original internal kinematics of the system at 
pericentre, removing a significant part of the rotation signal. 
This reduction is generally reinforced in successive passages. 
We observe that, in some cases, while the dwarf approaches the 
pericentre of its orbit there is a temporary increase of $|V_T|$ 
followed by a quick drop once pericentre is reached (see e.g., \texttt{subhaloID} = 428192). 
However, the fact that $|V_T|$ is measured in a perpendicular 
direction to the host makes it unlikely that these temporary 
increases are induced by tidal interaction.

The evolution of $A_{\rm grad}^{v_z}$ shows some similarities with 
$|V_T|$. The initially larger $A_{\rm grad}^{v_z}$ tends to be 
reduced by the successive pericentre passages. $A_{\rm grad}^{v_z}$ 
usually shows temporary increases asociated with pericentric 
passages, i.e.\ it increases as dwarfs approach their pericentres 
and drops as they abandon them. However, unlike $|V_T|$, these 
temporary increases in $A_{\rm grad}^{v_z}$ are generally observed 
during the majority of passages. 
These differences in the evolution of $|V_T|$ and $A_{\rm grad}^{v_z}$ 
are related to the way the host interacts with the satellite. $|V_T|$ 
and $A_{\rm grad}^{v_z}$ trace the presence of rotation in a satellite 
from an observational point of view, from the centre of the host. 
$|V_T|$ detects the presence of rotation as a 2D coherent motion in 
the plane of the sky and $A_{\rm grad}^{v_z}$ as a velocity gradient 
along the line of sight. $|V_T|$ and $A_{\rm grad}^{v_z}$ detect the 
presence of rotation in a satellite in perpendicular directions. 
The interaction of the host with the satellite takes place in a radial 
direction, i.e.\ along the $z$-axis of our reference frame. Therefore, 
this interaction is aligned with the direction in which we measure 
$A_{\rm grad}^{v_z}$ and is perpendicular to $|V_T|$. This makes 
$A_{\rm grad}^{v_z}$ more sensitive to such interactions with the host. 
During pericentric passages, the interaction reaches its 
highest value. For a satellite approaching pericentre for 
the first time, we observe how generally $A_{\rm grad}^{v_z}$ increases 
owing to the torque induced by the tidal force. Once the galaxy 
abandons the pericentre, both $|V_T|$ and $A_{\rm grad}^{v_z}$ decrease 
because their stellar content is stirred and a significant part of 
the rotation is removed. During successive passages, we are  
likely to detect temporary increases in  $A_{\rm grad}^{v_z}$ during 
pericentric passages because the direction of the tidal interaction 
is aligned with $v_z$. The pericentres, however, are less 
likely to induce temporary increases in $|V_T|$ given that the latter is 
in a direction perpendicular to the interaction and detects the 
presence of rotation as a 2D kinematic pattern. For both $|V_T|$ 
and $A_{\rm grad}^{v_z}$, we find that subsequent passages  tend 
to reduce the rotation of the system progressively, until reaching 
the low values that we detect at $z = 0$. In Table~\ref{tab:medians_z_0} 
we show the medians $|V_T|$ and $A_{\rm grad}^{v_z}$ for the satellites 
at $z = 0$.   
This is in agreement with the progressive reduction in the ratios 
of systems with detected rotation in the plane of the sky and 
gradients in $v_z$ (Fig.~\ref{fig:evolution_fraction}). The fact 
that interaction with the host has a different effect on $|V_T|$ 
and $A_{\rm grad}^{v_z}$ could explain why the ratios of galaxies with 
$|V_T| \neq 0$ and $A_{\rm grad}^{v_z} \neq 0$ evolve at slightly different 
rates (Fig.~\ref{fig:evolution_fraction}), in addition to the different
 sensitivity of the methods (Section~\ref{sec:evointkin}). 
The 
fraction of dwarfs with $A_{\rm grad}^{v_z} \neq 0$ decreases more gently because 
$A_{\rm grad}^{v_z}$ tends to increase during 
successive pericentres. 
Therefore, the reduction of the fraction is slower, given that there is 
a continuous contribution of dwarfs that show $A_{\rm grad}^{v_z} \neq 0$ 
owing to the tidal interaction with the host.

Previous studies based on $N$-body simulations have analysed 
the evolution of the rotation in dSph satellites of MW analogues 
(\citealt{Kazantzidis2011, Lokas2010, Lokas2011, Lokas2014, 
Lokas2015}), reporting  similar kinds of behaviour in the evolution 
of their kinematics. 
We note that these studies follow the evolution of the rotation 
around the shortest principal axis of the satellites, whereas we 
follow it from an observational point of view, i.e.\ we rely on 
$|V_T|$ and $A_{\rm grad}^{v_z}$.  
They find that the rotation tends to decrease, especially at
first pericentre, and that successive passages further reduce it. 
However, sometimes it is reported that the reduction of the 
rotation with the pericentres is not monotonic and may increase 
after a  passage owing to the orientation of the system during 
pericentre. We also find this in the evolution of some dwarfs, as 
can be seen in Fig.~\ref{fig:evolution_rotation} for \texttt{subhaloID} 
= 428192, where  $A_{\rm grad}^{v_z}$ slightly increases after the first 
pericentre. As for the temporary increases that we detect mostly in 
$A_{\rm grad}^{v_z}$ during pericentric passages, the simulated dwarfs  also 
show multiple examples of increases in the rotation when they approach 
pericentre followed by quick drop when they abandon it. We find 
that the evolution of the internal kinematics of the dwarf satellites 
of TNG50 is in good agreement with previous studies based on $N$-body 
simulations.

\begin{table*}
    \centering
    \caption{Median $|V_T|$, $A_{\rm grad}^{v_z}$, and line-of-sight 
velocity dispersion at $z = 0$ for the different samples of satellites 
galaxies. Rows: (1) Sample of satellites, (2) median $|V_T|$ at 
$z = 0$, and (3) median $A_{\rm grad}^{v_z}$ at $z = 0$. $\dag$ Median 
values and the interval between the 0.16 and 0.84 quantiles of the 
median $|V_T|$ and $A_{\rm grad}^{v_z}$ of all the realizations of the 
MC simulation (Section~\ref{subsec:analogues_mw_dsphs}). ${\dag\dag}$ 
values taken from \citet{McConnachie2020a, MartinezGarcia2021, MartinezGarcia2022}.}
    \begin{tabular}{l|c|c|c}
    \hline
         Sample & $|V_T| (z = 0)$ & $A_{\rm grad}^{v_z} (z = 0) $ & $\sigma$ \\
                & [$\kms$] & [$\kms$kpc$^{-1}$] & [$\kms$]\\
         \hline
         \hline
        TNG50 satellites  & $0.96 \pm 0.05$ & $0.99 \pm 0.04$ & $ 14.23 \pm 0.15$\\
        MW dSph analogues & $1.08 \pm 0.07$ & $0.91 \pm 0.07$ & $10.43  \pm  0.15$\\
        MW dSph analogues (MC)$^{\dag}$ & $2.00^{+1.72}_{-0.90}$ & $1.45^{+1.01}_{-0.60}$ & $10.43  \pm  0.15$\\
        MW dSphs$^{\dag\dag}$ & $1.86 \pm 1.64$ & $2.03 \pm 0.75$ & $9.15 \pm 4.68$\\
    \end{tabular}
    \label{tab:medians_z_0}
\end{table*}

\subsection{Traces of interaction with the host in the internal kinematics}
\label{sec:col_map_rot_grad}

\begin{figure}
    \centering
    \includegraphics[width=\linewidth]{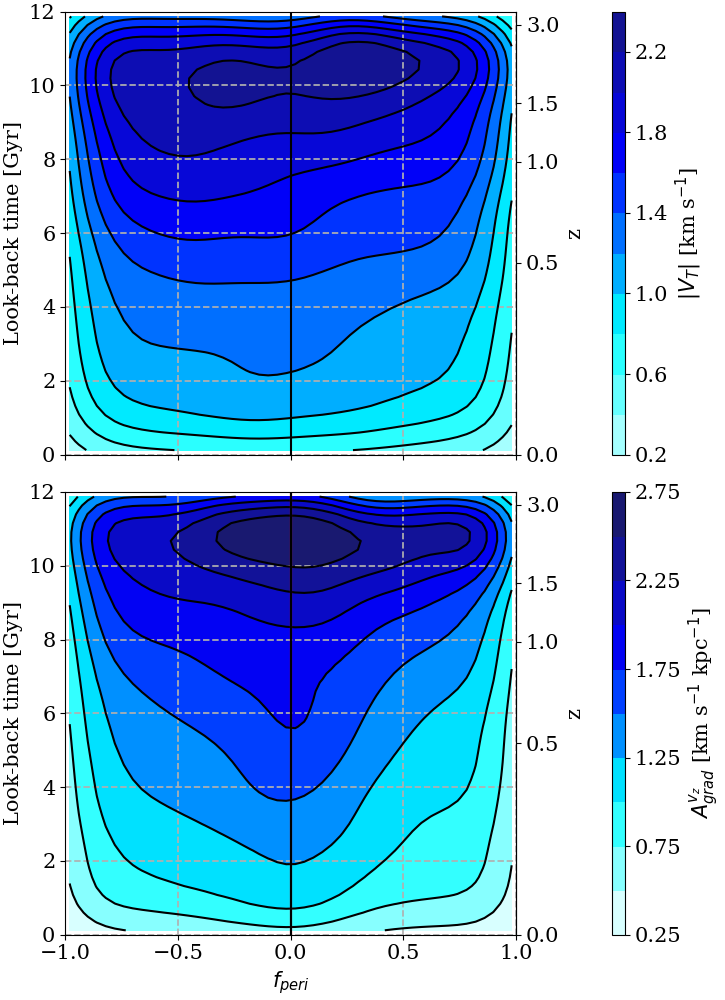}
    \caption{Evolution of the internal kinematics of the satellites 
as a function of the relative position  in their orbits. The panels 
show $f_{\mathrm{peri}}$ and look-back time for all the dwarfs and 
snapshots. Colour-filled contours represent areas with similar $|V_T|$ 
(top panel) and $A_{\rm grad}^{v_z}$ (bottom panel).}
    \label{fig:map_fperi}
\end{figure}

In order to get a more comprehensive picture of the evolution 
of the internal kinematics of the dwarf satellites and the role 
of the interaction with the host over time, we compared simultaneously the 
kinematic features with the relative position of all the dwarfs 
in their orbits during all the snapshots of the simulation. 
This allows us to have a more global view of the evolution of the 
kinematics  and to assess our findings of the previous section.
The relative position of a galaxy in its orbit is given by 
$f_{\mathrm{peri}}$, which is defined as follows: $f_{\mathrm{peri}} = 
\mathrm{sgn}(v_R^{\rm GC}) (d_{\rm GC} - d_p) / (d_a - d_p)$, where $d_{\rm GC}$ 
is the distance of the satellite to the host, $v_R^{\rm GC}$ is the radial 
velocity of the satellite with respect to the host, $d_a$ is the apocentric 
distance, and $d_p$ is the pericentric distance. $f_{\mathrm{peri}} = 0$ 
means that the satellite is at the pericentre of its orbit, whereas 
 $|f_{\mathrm{peri}}| = 1$ when it is at its apocentre. 
$f_{\mathrm{peri}} > 0$ implies that the galaxy is moving towards 
apocentre whereas $f_{\mathrm{peri}} < 0$ is for galaxies approaching 
pericentre. 
$f_{\mathrm{peri}}$ is similar to the true anomaly in Keplerian orbits
 and is a very useful metric, because it provides a clear idea of the
 position and direction of a satellite along its orbit. 
In Fig.~\ref{fig:map_fperi}, we represent 
$f_{\mathrm{peri}}$ and the look-back time for all the selected 
satellites and all the snapshots of the simulation. The data were 
binned and a Gaussian filter applied. Colour-filled contours encompass 
areas with similar $|V_T|$ (top panel) or $A_{\rm grad}^{v_z}$ (bottom panel). 
Figure~\ref{fig:map_fperi} allows the simultaneous study of all the dwarfs 
and snapshots, being a generalized version of Fig.~\ref{fig:evolution_rotation}, 
and providing additional confirmation to our findings regarding the 
evolution of the kinematics of individual satellites and their interaction 
with the host.

At first glance, we observe a clear decrease of both  $|V_T|$ and 
$A_{\rm grad}^{v_z}$ with time. Initially, the rotation velocity 
in the plane of the sky and the amplitude of the gradients in 
these systems were larger. There is a progressive reduction in 
both features until the low values that we currently
 observe at $z = 0$ are reached. This decrease is in agreement with the  
reduction of the fraction of systems with detected rotation 
in the plane of the sky and gradients in $v_z$ of 
Fig.~\ref{fig:evolution_fraction} and the progressive reduction 
of  $|V_T|$ and $A_{\rm grad}^{v_z}$ with time of Figure~\ref{fig:evolution_rotation}. 
Figure~\ref{fig:map_fperi} shows that at a given look-back time, 
$|V_T|$ and $A_{\rm grad}^{v_z}$ tend to be systematically slightly 
larger in the region where $f_{\mathrm{peri}} < 0$, i.e.\ for 
galaxies approaching their pericentres, compared to 
$f_{\mathrm{peri}} > 0$, i.e.\ for galaxies abandoning it. This 
is consistent with the progressive reduction of $|V_T|$ and 
$A_{\rm grad}^{v_z}$ associated with the pericentric passages observed 
in the analysis of the individual satellites.
However, we observe significant differences in the evolution 
of $|V_T|$ and $A_{\rm grad}^{v_z}$. During the whole simulation, 
for a given snapshot  $A_{\rm grad}^{v_z}$ tends to reach its 
highest value in the vicinity of pericentre. This is 
in agreement with the evolution of $A_{\rm grad}^{v_z}$ observed 
in individual satellites: $A_{\rm grad}^{v_z}$ tends to increase 
as the satellites approach their pericentre, where 
$A_{\rm grad}^{v_z}$ reaches its higher value and is then reduced 
as the satellites move away from pericentre. For $|V_T|$
 we find a more monotonic evolution, with a progressive 
reduction of $|V_T|$ and without higher values around 
petricentre. This is consistent with  the analysis of individual 
satellites from which, generally, we do not find temporary increases 
of $|V_T|$ during pericentric passages.
In conclusion, the different way in which $|V_T|$ and $A_{\rm grad}^{v_z}$ 
are affected by the interaction with the host (Section~\ref{sec:evo_v_3}) 
explains the different evolution of these features with time.

\subsection{Analogues of the observed MW dSph satellites}
\label{subsec:analogues_mw_dsphs}

Previous observational studies have derived $V_T$ and $A_{\rm grad}^{v_z}$ 
for dSph satellites of the MW (\citealt{MartinezGarcia2021, MartinezGarcia2022}).
In order 
to compare the internal kinematics of real and simulated 
satellites, we further restrict our sample to match the 
properties of the observed dSphs. 
We select a subsample of TNG50 dSph satellites whose stellar
 mass, line-of-sight velocity dispersion, and $M_V$ are compatible
 with those of Carina, Draco, Fornax, Sculptor, Sextans, and 
Ursa Minor (\citealt{McConnachie2012}). The range of these 
properties is represented as shaded green areas in 
Fig.~\ref{fig:properties}, where it can be seen that they represent 
the faint and low-mass end of the corresponding distributions. The 
resulting subsample consists of 251 dSph satellites, hereafter 
referred to as `MW dSph analogues'. The internal  kinematics of 
the MW dSph analogues show a behaviour similar to  the other
simulated galaxies. 

We find that the fraction of systems rotating in the plane 
of the sky in the subsample of MW dSph analogues was initially 
similar to the other simulated satellites ($\sim$65\%), 
and decreased until reaching 39\% at $z = 0$. 
This ratio is similar (45\%) to that for the TNG50 dSphs at 
$z = 0$. 
We find similar results for the fraction of MW dSph analogues with gradients in 
$v_z$. Initially, the fraction
 was similar to the other TNG50 satellites ($\sim$80\%) 
and decreased to 56\% at $z = 0$. The ratio of MW dSph 
analogues with gradients in $v_z$ at $z = 0$ is similar (65\%) to 
that for the TNG50 dSphs at $z = 0$.
We observe that the behaviour in the reduction of the 
ratios for the full sample of TNG50 satellites and the 
subsample of MW dSph analogues is similar. If anything, 
the differences in the ratios at $z = 0$ could be related 
to the lower masses of the MW dSph analogues compared to 
the selected satellites of TNG50. These lower masses make
it more difficult to detect rotation in the plane of the 
sky and gradients in $v_z$, and also would make the MW 
dSph analogues more sensitive to interaction with the 
host so that they further reduced their rotation. 
The final ratios suggest that the presence of detectable 
rotation in satellites of MW analogues is not uncommon. For 
the MW satellites, the presence of rotation in the plane of 
the sky has been reported for Carina, Fornax, Sagittarius,
 and Sculptor, and gradients in $v_z$ for Carina, Draco, 
Fornax, Sagittarius, and Ursa Minor (\citealt{delPino2021, 
MartinezGarcia2021, MartinezGarcia2022}). Thus far, the 
number of MW satellites for which deriving the internal 
kinematics is possible is small and hence the number of dSph 
satellites for which coherent motions are detected. The 
combination of the lack of gas in dSphs, their distance, and 
technical limitations make it difficult to study their internal 
kinematics. However, future measurements of PMs (e.g.\ {\itshape Gaia} Data 
Release 4) and $\vlos$ may allow us to study the internal 
kinematics of more satellites of the MW.

\begin{figure}
    \centering
    \includegraphics[width=\linewidth]{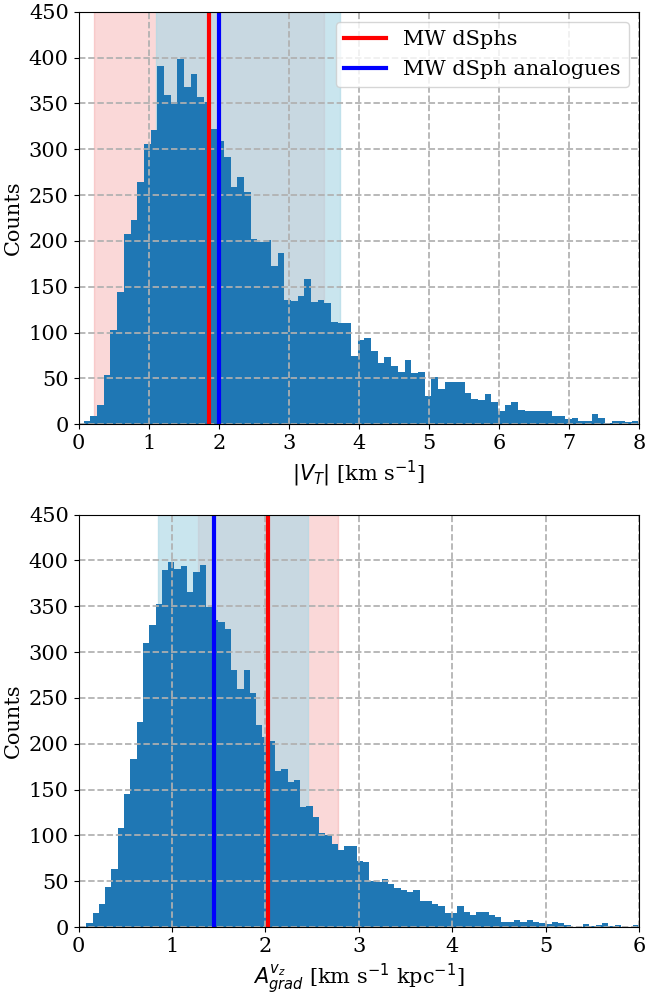}
    \caption{Distribution of the median $|V_T|$ (top panel) 
and  the median $A_{\rm grad}^{v_z}$ (bottom panel) of the output 
of the MC simulation of Section ~\ref{subsec:analogues_mw_dsphs}. 
    The blue vertical lines and blue-shaded areas represent the 
median values and the interval between the 0.16 and 0.84 quantiles 
of the medians $|V_T|$ and $A_{\rm grad}^{v_z}$ for all the realizations. 
    The red vertical lines and the red-shaded areas represent the 
median and its standard error for $|V_T|$ and $A_{\rm grad}^{v_z}$ 
for the observed MW dSphs (\citealt{MartinezGarcia2021, MartinezGarcia2022}).}
    \label{fig:distr_feat}
\end{figure}

The evolution of the internal kinematics of satellites 
in the subsample of MW dSph analogues reveals a similar 
behaviour to the other TNG50 dwarf satellites. This can 
be seen for individual satellites in Fig.~\ref{fig:evolution_rotation},
 where the panels of \texttt{subhaloID} 554807 and 516783 
correspond to dwarfs in the subsample.
The MW dSph analogues had initially higher values of 
$|V_T|$ and $A_{\rm grad}^{v_z}$ that progressively decreased 
with time, as we find for the TNG50 satellites. We also 
observe temporary increases in $A_{\rm grad}^{v_z}$ with the 
pericentre passages, as discussed in Section~\ref{sec:evo_v_3}.
The  medians of $|V_T|$ and $A_{\rm grad}^{v_z}$ for all the 
dwarfs in the subsample at $z = 0$ are $1.08 \pm 0.07 \kms$ 
and  $0.91 \pm 0.07$ km s$^{-1}$ kpc$^{-1}$ respectively. 
These values can be found in Table~\ref{tab:medians_z_0},  
where it can be seen that they are similar to the other 
satellites of TNG50; nonetheless they are lower than the 
median $|V_T|$ and $A_{\rm grad}^{v_z}$ for the MW dSph satellites 
(\citealt{MartinezGarcia2021, MartinezGarcia2022}).
These discrepancies could be explained by the greater 
uncertainties of $|V_T|$ and $A_{\rm grad}^{v_z}$ in 
observational studies, since these take into account more 
sources of error, such as the uncertainties in PMs and $\vlos$ 
for the bulk and individual stars, among others (for more 
details see section 2.3 of \citealt{MartinezGarcia2021} and 
section 2.2 of \citealt{MartinezGarcia2022}). 
For comparison between observed and simulated MW satellites,
 we used a Monte Carlo (MC) scheme. In each iteration we 
randomly choose six MW dSph analogues in order to have a sample 
whose size is consistent with the number of MW dSphs for which 
their internal kinematics have been derived 
(\citealt{MartinezGarcia2021, MartinezGarcia2022}). For each 
of the sampled galaxies, we add to its $V_T$ a random number 
from a normal distribution centred on zero and with a dispersion 
equal to the observational error of $V_T$ of one of the observed 
MW dSph satellites of \citet{MartinezGarcia2021} chosen at 
random. We then calculate the median value of $|V_T|$ of the 
six sampled galaxies. We proceed similarly for $A_{\rm grad}^{v_z}$, 
adding a random number from a normal distribution centred on
zero and a dispersion equal to the observational error of 
$A_{\rm grad}^{v_z}$ of one of the observed MW dSphs of \citet{MartinezGarcia2022} 
chosen at random. We then calculate the median value of 
$A_{\rm grad}^{v_z}$ of the six sampled galaxies. This procedure 
is repeated for $10^4$ iterations. The distributions of the
 medians of $|V_T|$ and $A_{\rm grad}^{v_z}$ for all the iterations 
are shown in Figure~\ref{fig:distr_feat}. 
We include blue vertical lines and blue-shaded areas that 
represent the median and  the interval between the 0.16 and 
0.84 quantiles of all the realizations for $|V_T|$ 
($2.00^{+1.72}_{-0.90}\kms$) and $A_{\rm grad}^{v_z}$ 
($1.45^{+1.01}_{-0.60}$ $\kms$kpc$^{-1}$). These values 
can be found in Table~\ref{tab:medians_z_0}. We also 
include in Figure~\ref{fig:distr_feat} red vertical lines 
and red-shaded areas that represent the median and its 
standard error for $|V_T|$ and $A_{\rm grad}^{v_z}$ for the MW 
dSph satellites (\citealt{MartinezGarcia2021, MartinezGarcia2022}). 
We observe that the median values of the kinematic features 
obtained through the MC simulation and the median values for 
the observed MW dSphs are in good agreement. In the case of 
$|V_T|$, observations and simulations show very similar values. 
For $A_{\rm grad}^{v_z}$, the results are also similar, but not as 
close as in the case of $|V_T|$. We note that for both features 
the median values for the observed and simulated MW dSphs are 
consistent at the $1\sigma$ level.

\begin{figure}
    \centering
    \includegraphics[width=\linewidth]{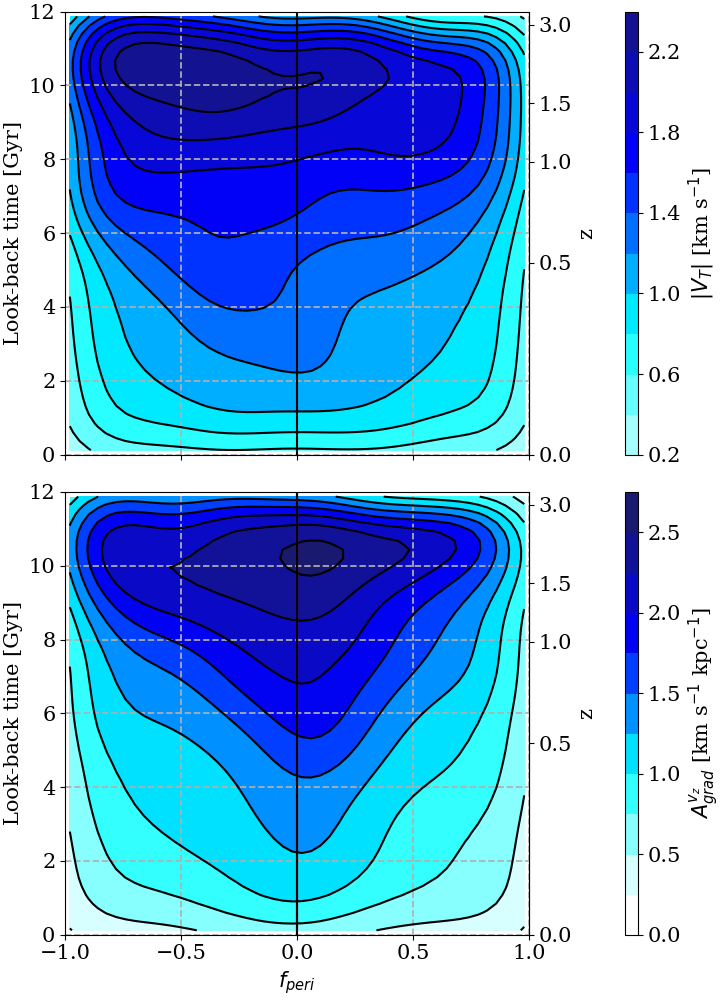}
    \caption{Evolution of the internal kinematics of the 
satellites as a function of the relative position  in their 
orbits for the subsample of MW dSph analogues. Markers 
coincide with those of Figure~\ref{fig:map_fperi}.}
    \label{fig:map_fperi_MW}
\end{figure}

\begin{figure}
    \centering
    \includegraphics[width=\linewidth]{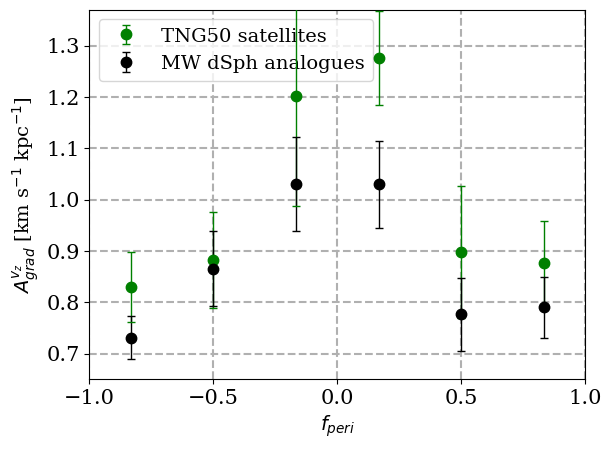}
    \caption{Amplitude of the gradients in $v_z$ of the 
satellites and their relative position in their orbits  
during the last 0.5 Gyr. The points represent the median amplitude 
of the gradients in $v_z$ of satellites in evenly spaced bins 
in $f_{\rm peri}$. Green points  represent all the  selected satellites 
from TNG50 and black points represent the subsample of MW dSph analogues. }
    \label{fig:f_peri_z_0_comb}
\end{figure}

We analysed the impact of the interaction with the host on the 
internal kinematics of the MW dSph analogues. In 
Fig.~\ref{fig:map_fperi_MW} we show $f_{\mathrm{peri}}$ and the 
look-back time for all the  MW dSph analogues and all the 
snapshots of the simulation. We include colour-filled contours of 
$|V_T|$ (top panel) and $A_{\rm grad}^{v_z}$ (bottom panel). 
Figure~\ref{fig:map_fperi_MW} is an analogue version of  
Fig.~\ref{fig:map_fperi}, restricted to the MW dSph analogues. 
We observe similar trends in both figures, yet for the MW dSph 
analogues the plot is noisier owing to the lower number of 
 galaxies analysed. 
The trends in Fig.~\ref{fig:map_fperi_MW} are sharper compared 
to Fig.~\ref{fig:map_fperi} an effect that is likely to be caused by the 
lower masses of the MW dSph analogues, which makes them more 
susceptible to the influence of the host. 
In order to compare these trends with previous observational 
studies, in Fig.~\ref{fig:f_peri_z_0_comb} we show 
$f_{\mathrm{peri}}$ and $A_{\rm grad}^{v_z}$ for all the TNG50 
satellites and the MW dSph analogues at look-back time $< 0.5$ 
Gyr. We cannot restrict the samples strictly to $z = 0$ because 
the sample of MW dSph analogues contains a low number of 
galaxies. Including the last 0.5 Gyr allows us to increase 
the statistics while analysing a time span close enough to $z = 0$. 
We used evenly spaced bins along $f_{\mathrm{peri}}$ and 
calculated the median $A_{\rm grad}^{v_z}$ in each bin. 
Figure~\ref{fig:f_peri_z_0_comb} represents an analogous version of 
Fig.~\ref{fig:map_fperi} and Fig.~\ref{fig:map_fperi_MW} at 
$z \sim 0$.  Both the whole sample of TNG50 satellites and the 
subsample of MW dSph analogues show a similar behaviour, 
consistent with Fig.~\ref{fig:map_fperi} and Fig.~\ref{fig:map_fperi_MW},  
and match the reported trend in $A_{\rm grad}^{v_z}$ for the MW 
dSph satellites: the amplitude of $A_{\rm grad}^{v_z}$ tends to 
increase as the MW dSphs  approach their pericentres  and 
$A_{\rm grad}^{v_z}$  tends to decrease for dwarfs as they head 
towards their apocentres, the largest values of 
$A_{\rm grad}^{v_z}$  being reached in the vicinity of the pericentre (see top 
panel of fig. 8 of \citealt{MartinezGarcia2022}). 
The fact that simulated and observed galaxies show a similar 
trend, in which the amplitude of their gradients depends on the 
position in their orbits suggests that the gradients of the MW 
dSph satellites are caused by interaction with the MW, and thus 
the interaction with the host does have an impact on the evolution 
of the internal kinematics of its satellites.

\section{Conclusions}
\label{sec:conclusions}
In this paper, we have studied the internal kinematics of dwarf 
satellites of the TNG50 simulation. We looked for dwarfs 
resembling those of the LG, so we explored MW/M31-like hosts and 
analysed their subhaloes. We selected 1017 dwarf satellites for 
which we derived their internal kinematics in every snapshot. 
This allowed us to obtain a historic record of their rotation 
velocity in the plane of the sky ($V_T$) and the amplitude of their velocity
gradients along the line of sight ($A_{\rm grad}^{v_z}$) corrected for 
perspective effects.

For the vast majority of the systems, we initially detect  
rotation in the plane of the sky ($\sim$65\%) or velocity 
gradients along the line-of-sight ($\sim$80\%). 
These figures were progressively reduced until reaching 45\% 
and 68\%, respectively, at $z = 0$. Therefore, this suggests 
that the presence of detectable rotation and  gradients along 
the line of sight in satellites of MW/M31-like hosts is not an 
uncommon feature at present. We studied possible differences in 
the evolution of the internal kinematics between dSphs and the 
other dwarfs (non-dSphs, for simplicity). We observe that 
the fraction of systems with detected rotation in the
 plane of the sky and the velocity gradients was originally  similar, 
regardless of the type of dwarf. However, around look-back time 
$\sim$10 Gyr the trends started diverging. We observe a sooner 
and faster decrease of the ratios for the dSphs.  At $z = 0$,  
ratios are lower for dSphs. 
Additionally,  we find that $|V_T|$  and $A_{\rm grad}^{v_z}$ have 
different distributions at $z = 0$ for dSphs and non-dSphs. 
These differences are likely to be due to the different orbital 
histories of different types of dwarfs. We find that dSphs generally 
perform their first pericentric passage sooner and  experience 
more passages than non-dSphs, which explains the earlier and 
more intense decreases in the ratio of rotating systems in the 
plane of the sky and the ratio of satellites with velocity 
gradients, and thus the differences in the distributions.

We find that interaction with the host has impacted the 
evolution of internal kinematics of the satellites. 
Our analysis of the evolution of $|V_T|$ in individual 
satellites reveals that, initially, the satellites had larger 
values of $|V_T|$ that were progressively reduced. 
After first  pericentre there is a significant reduction 
of the rotation in the plane of the sky. The tidal 
interaction during the passage disrupts the original 
kinematics, thereby severely reducing the rotation in the plane
 of the sky. The average rotation velocity is further reduced 
during successive pericentric passages. 
Previous studies based on $N$-body simulations have found 
a similar evolution in the rotation velocity of the dSphs 
as they interact with their host galaxy
 (\citealt{Kazantzidis2011, Lokas2010, Lokas2011, Lokas2014, Lokas2015}).
The evolution of $A_{\rm grad}^{v_z}$ is similar, the 
originally larger values of $A_{\rm grad}^{v_z}$  are 
progressively reduced during pericentric passages. 
We observe temporary increases of $A_{\rm grad}^{v_z}$ during 
pericentric passages, because the interaction with 
the host takes place in the direction in which the velocity 
gradients are measured.  Thus, during pericentre the 
tidal forces tend to induce these velocity gradients in the 
satellites. 

We then analysed the internal kinematics of all the satellites 
and snapshots simultaneously. We detect a general reduction 
of $|V_T|$ and $A_{\rm grad}^{v_z}$ with time. We also find that 
$|V_T|$ and $A_{\rm grad}^{v_z}$ show a different evolution. For 
$|V_T|$ we observe a monotonic reduction; however, for 
$A_{\rm grad}^{v_z}$ we observe that during the whole simulation, 
at any given snapshot $A_{\rm grad}^{v_z}$ reaches its highest 
value close  to pericentre. These trends are in agreement 
with those that we observed in the internal kinematics of 
individual satellites of TNG50, and are consistent with the 
different ways in which $|V_T|$ and $A_{\rm grad}^{v_z}$ are impacted 
by interaction with the host.

We finally restricted our sample of simulated satellites 
to those that match the properties of the observed MW dSphs, 
finding a similar evolution of the internal kinematics 
compared to the entire sample of TNG50 satellites. We observe 
that the trends in  the evolution of $|V_T|$ and $A_{\rm grad}^{v_z}$ 
are sharper for the MW dSph analogues. This could be because 
they have lower masses than the other TNG50 
satellites and are  thus more easily affected by interaction 
with the host. The trend in $A_{\rm grad}^{v_z}$ for MW dSph 
analogues and TNG50 satellites is similar to that reported 
for the actual MW dSph satellites,  
for which  $A_{\rm grad}^{v_z}$ tends to increase as the satellites 
move towards their pericentres and is reduced as  dwarfs move 
towards their apocentres, with $A_{\rm grad}^{v_z}$ reaching its
 maximum value in the vicinity of pericentre  (\citealt{MartinezGarcia2022}). 
The presence of this trend in both simulated and actual dSph 
satellites suggests that interaction with the MW is 
responsible for the observed velocity gradients, and that 
interaction with the host affects the evolution of the internal 
kinematics of its satellites.

\section*{Acknowledgements}
We acknowledge support from the Spanish Agencia Estatal de 
Investigación del Ministerio de Ciencia e Innovacion (AEI-MICINN) 
under grant `Proyectos de I+D+i' with references AYA2017-89841-P 
and PID2020-115981GB-I00. AdP acknowledges the financial 
support from the European Union - NextGenerationEU and the 
Spanish Ministry of Science and Innovation through the Recovery 
and Resilience Facility project J-CAVA. The authors also acknowledge 
all the open-source software involved in this study, especially 
TOPCAT, Python, and git.

%%%%%%%%%%%%%%%%%%%%%%%%%%%%%%%%%%%%%%%%%%%%%%%%%%
\section*{Data Availability}
All the data underlying this article are publicly available. 
The IllustrisTNG simulations are publicly available and can be 
found in \url{www.tng-project.org/data}.



\bibliographystyle{mnras}
\bibliography{biblio.bib} % if your bibtex file is called example.bib

\begin{thebibliography}{}
\makeatletter
\relax
\def\mn@urlcharsother{\let\do\@makeother \do\$\do\&\do\#\do\^\do\_\do\%\do\~}
\def\mn@doi{\begingroup\mn@urlcharsother \@ifnextchar [ {\mn@doi@} {\mn@doi@[]}}
\def\mn@doi@[#1]#2{\def\@tempa{#1}\ifx\@tempa\@empty \href {http://dx.doi.org/#2} {doi:#2}\else \href {http://dx.doi.org/#2} {#1}\fi \endgroup}
\def\mn@eprint#1#2{\mn@eprint@#1:#2::\@nil}
\def\mn@eprint@arXiv#1{\href {http://arxiv.org/abs/#1} {{\tt arXiv:#1}}}
\def\mn@eprint@dblp#1{\href {http://dblp.uni-trier.de/rec/bibtex/#1.xml} {dblp:#1}}
\def\mn@eprint@#1:#2:#3:#4\@nil{\def\@tempa {#1}\def\@tempb {#2}\def\@tempc {#3}\ifx \@tempc \@empty \let \@tempc \@tempb \let \@tempb \@tempa \fi \ifx \@tempb \@empty \def\@tempb {arXiv}\fi \@ifundefined {mn@eprint@\@tempb}{\@tempb:\@tempc}{\expandafter \expandafter \csname mn@eprint@\@tempb\endcsname \expandafter{\@tempc}}}

\bibitem[\protect\citeauthoryear{Amorisco \& Evans}{Amorisco \& Evans}{2012}]{Amorisco2012}
Amorisco N.~C.,  Evans N.~W.,  2012, \mn@doi [\mnras] {10.1111/j.1365-2966.2011.19684.x}, 419, 184

\bibitem[\protect\citeauthoryear{Battaglia, Helmi, Tolstoy, Irwin, Hill  \& Jablonka}{Battaglia et~al.}{2008}]{Battaglia2008b}
Battaglia G.,  Helmi A.,  Tolstoy E.,  Irwin M.,  Hill V.,   Jablonka P.,  2008, \mn@doi [\apj] {10.1086/590179}, 681, L13

\bibitem[\protect\citeauthoryear{Battaglia, Tolstoy, Helmi, Irwin, Parisi, Hill  \& Jablonka}{Battaglia et~al.}{2011}]{Battaglia2011}
Battaglia G.,  Tolstoy E.,  Helmi A.,  Irwin M.,  Parisi P.,  Hill V.,   Jablonka P.,  2011, \mn@doi [\mnras] {10.1111/j.1365-2966.2010.17745.x}, 411, 1013

\bibitem[\protect\citeauthoryear{Blumenthal, Faber, Primack  \& Rees}{Blumenthal et~al.}{1984}]{Blumenthal1984}
Blumenthal G.~R.,  Faber S.~M.,  Primack J.~R.,   Rees M.~J.,  1984, \mn@doi [Nature] {10.1038/311517a0}, 311, 517

\bibitem[\protect\citeauthoryear{Bullock \& Boylan-Kolchin}{Bullock \& Boylan-Kolchin}{2017}]{Bullock2017}
Bullock J.~S.,  Boylan-Kolchin M.,  2017, \mn@doi [Annual Review of Astronomy and Astrophysics] {10.1146/annurev-astro-091916-055313}, 55, 343

\bibitem[\protect\citeauthoryear{Cappellari \& Copin}{Cappellari \& Copin}{2003}]{Capellari2003}
Cappellari M.,  Copin Y.,  2003, \mn@doi [\mnras] {10.1046/j.1365-8711.2003.06541.x}, 342, 345

\bibitem[\protect\citeauthoryear{{Cardona-Barrero}, {Battaglia}, {Di Cintio}, {Revaz}  \& {Jablonka}}{{Cardona-Barrero} et~al.}{2021}]{CardonaBarrero2021}
{Cardona-Barrero} S.,  {Battaglia} G.,  {Di Cintio} A.,  {Revaz} Y.,   {Jablonka} P.,  2021, \mn@doi [\mnras] {10.1093/mnrasl/slab059}, \href {https://ui.adsabs.harvard.edu/abs/2021MNRAS.505L.100C} {505, L100}

\bibitem[\protect\citeauthoryear{{Davis}, {Efstathiou}, {Frenk}  \& {White}}{{Davis} et~al.}{1985}]{Davis1985}
{Davis} M.,  {Efstathiou} G.,  {Frenk} C.~S.,   {White} S.~D.~M.,  1985, \mn@doi [\apj] {10.1086/163168}, \href {https://ui.adsabs.harvard.edu/abs/1985ApJ...292..371D} {292, 371}

\bibitem[\protect\citeauthoryear{Dekel \& Silk}{Dekel \& Silk}{1986}]{Dekel1986}
Dekel A.,  Silk J.,  1986, \mn@doi [ApJ] {10.1086/164050}, 303, 39

\bibitem[\protect\citeauthoryear{{\VAN{Del Pino}{del }{del }{Pino}}, Aparicio, Hidalgo  \& {\L}okas}{{\VAN{Del Pino}{del }{del }{Pino}} et~al.}{2017}]{delPino2017}
{\VAN{Del Pino}{del }{del }{Pino}} A.,  Aparicio A.,  Hidalgo S.~L.,   {\L}okas E.~L.,  2017, \mn@doi [\mnras] {10.1093/mnras/stw3016}, 465, 3708

\bibitem[\protect\citeauthoryear{{\VAN{Del Pino}{del }{del }{Pino}}, Fardal, van~der Marel, {\L}okas, Mateu  \& Sohn}{{\VAN{Del Pino}{del }{del }{Pino}} et~al.}{2021}]{delPino2021}
{\VAN{Del Pino}{del }{del }{Pino}} A.,  Fardal M.~A.,  van~der Marel R.~P.,  {\L}okas E.~L.,  Mateu C.,   Sohn S.~T.,  2021, \mn@doi [\apj] {10.3847/1538-4357/abd5bf}, 908, 244

\bibitem[\protect\citeauthoryear{Dolag, Borgani, Murante  \& Springel}{Dolag et~al.}{2009}]{Dolag2009}
Dolag K.,  Borgani S.,  Murante G.,   Springel V.,  2009, \mn@doi [\mnras] {10.1111/j.1365-2966.2009.15034.x}, 399, 497

\bibitem[\protect\citeauthoryear{{Ebrov{\'a}} \& {{\L}okas}}{{Ebrov{\'a}} \& {{\L}okas}}{2015}]{Ebrova2015}
{Ebrov{\'a}} I.,  {{\L}okas} E.~L.,  2015, \mn@doi [\apj] {10.1088/0004-637X/813/1/10}, \href {https://ui.adsabs.harvard.edu/abs/2015ApJ...813...10E} {813, 10}

\bibitem[\protect\citeauthoryear{{Engler} et~al.,}{{Engler} et~al.}{2021}]{Engler2021}
{Engler} C.,  et~al., 2021, \mn@doi [\mnras] {10.1093/mnras/stab2437}, \href {https://ui.adsabs.harvard.edu/abs/2021MNRAS.507.4211E} {507, 4211}

\bibitem[\protect\citeauthoryear{{Engler}, {Pillepich}, {Joshi}, {Pasquali}, {Nelson}  \& {Grebel}}{{Engler} et~al.}{2022}]{Engler2022}
{Engler} C.,  {Pillepich} A.,  {Joshi} G.~D.,  {Pasquali} A.,  {Nelson} D.,   {Grebel} E.~K.,  2022, \mn@doi [arXiv e-prints] {10.48550/arXiv.2211.00010}, \href {https://ui.adsabs.harvard.edu/abs/2022arXiv221100010E} {p. arXiv:2211.00010}

\bibitem[\protect\citeauthoryear{Fabrizio et~al.,}{Fabrizio et~al.}{2016}]{Fabrizio2016}
Fabrizio M.,  et~al., 2016, \mn@doi [\apj] {10.3847/0004-637x/830/2/126}, 830, 126

\bibitem[\protect\citeauthoryear{{Feast}, {Thackeray}  \& {Wesselink}}{{Feast} et~al.}{1961}]{Feast1961}
{Feast} M.~W.,  {Thackeray} A.~D.,   {Wesselink} A.~J.,  1961, \mn@doi [\mnras] {10.1093/mnras/122.5.433}, \href {https://ui.adsabs.harvard.edu/abs/1961MNRAS.122..433F} {122, 433}

\bibitem[\protect\citeauthoryear{{Gaia Collaboration} et~al.,}{{Gaia Collaboration} et~al.}{2016}]{GaiaCollaboration2016}
{Gaia Collaboration} et~al., 2016, \mn@doi [\aap] {10.1051/0004-6361/201629272}, 595, A1

\bibitem[\protect\citeauthoryear{Irwin \& Hatzidimitriou}{Irwin \& Hatzidimitriou}{1995}]{Irwin1995}
Irwin M.,  Hatzidimitriou D.,  1995, \mn@doi [\mnras] {10.1093/mnras/277.4.1354}, 277, 1354

\bibitem[\protect\citeauthoryear{Joshi, Pillepich, Nelson, Zinger, Marinacci, Springel, Vogelsberger  \& Hernquist}{Joshi et~al.}{2021}]{Joshi2021}
Joshi G.~D.,  Pillepich A.,  Nelson D.,  Zinger E.,  Marinacci F.,  Springel V.,  Vogelsberger M.,   Hernquist L.,  2021, \mn@doi [\mnras] {10.1093/mnras/stab2573}, 508, 1652

\bibitem[\protect\citeauthoryear{Kazantzidis, {\L}okas, Callegari, Mayer  \& Moustakas}{Kazantzidis et~al.}{2011}]{Kazantzidis2011}
Kazantzidis S.,  {\L}okas E.~L.,  Callegari S.,  Mayer L.,   Moustakas L.~A.,  2011, \mn@doi [\apj] {10.1088/0004-637x/726/2/98}, 726, 98

\bibitem[\protect\citeauthoryear{Kirby, Bullock, Boylan-Kolchin, Kaplinghat  \& Cohen}{Kirby et~al.}{2014}]{Kirby2014}
Kirby E.~N.,  Bullock J.~S.,  Boylan-Kolchin M.,  Kaplinghat M.,   Cohen J.~G.,  2014, \mn@doi [\mnras] {10.1093/mnras/stu025}, 439, 1015

\bibitem[\protect\citeauthoryear{Kleyna, Wilkinson, Evans, Gilmore  \& Frayn}{Kleyna et~al.}{2002}]{Kleyna2002}
Kleyna J.,  Wilkinson M.~I.,  Evans N.~W.,  Gilmore G.,   Frayn C.,  2002, \mn@doi [\mnras] {10.1046/j.1365-8711.2002.05155.x}, 330, 792

\bibitem[\protect\citeauthoryear{{Klimentowski}, {{\L}okas}, {Kazantzidis}, {Mayer}  \& {Mamon}}{{Klimentowski} et~al.}{2009}]{Klimentowski2009}
{Klimentowski} J.,  {{\L}okas} E.~L.,  {Kazantzidis} S.,  {Mayer} L.,   {Mamon} G.~A.,  2009, \mn@doi [\mnras] {10.1111/j.1365-2966.2009.15046.x}, \href {https://ui.adsabs.harvard.edu/abs/2009MNRAS.397.2015K} {397, 2015}

\bibitem[\protect\citeauthoryear{Koch, Kleyna, Wilkinson, Grebel, Gilmore, Evans, Wyse  \& Harbeck}{Koch et~al.}{2007a}]{Koch2007b}
Koch A.,  Kleyna J.~T.,  Wilkinson M.~I.,  Grebel E.~K.,  Gilmore G.~F.,  Evans N.~W.,  Wyse R. F.~G.,   Harbeck D.~R.,  2007a, \mn@doi [\aj] {10.1086/519380}, 134, 566

\bibitem[\protect\citeauthoryear{Koch, Wilkinson, Kleyna, Gilmore, Grebel, Mackey, Evans  \& Wyse}{Koch et~al.}{2007b}]{Koch2007a}
Koch A.,  Wilkinson M.~I.,  Kleyna J.~T.,  Gilmore G.~F.,  Grebel E.~K.,  Mackey A.~D.,  Evans N.~W.,   Wyse R. F.~G.,  2007b, \mn@doi [\apj] {10.1086/510879}, 657, 241

\bibitem[\protect\citeauthoryear{{{\L}okas}, {Kazantzidis}, {Majewski}, {Law}, {Mayer}  \& {Frinchaboy}}{{{\L}okas} et~al.}{2010}]{Lokas2010}
{{\L}okas} E.~L.,  {Kazantzidis} S.,  {Majewski} S.~R.,  {Law} D.~R.,  {Mayer} L.,   {Frinchaboy} P.~M.,  2010, \mn@doi [\apj] {10.1088/0004-637X/725/2/1516}, \href {https://ui.adsabs.harvard.edu/abs/2010ApJ...725.1516L} {725, 1516}

\bibitem[\protect\citeauthoryear{{{\L}okas}, {Kazantzidis}  \& {Mayer}}{{{\L}okas} et~al.}{2011}]{Lokas2011}
{{\L}okas} E.~L.,  {Kazantzidis} S.,   {Mayer} L.,  2011, \mn@doi [\apj] {10.1088/0004-637X/739/1/46}, \href {https://ui.adsabs.harvard.edu/abs/2011ApJ...739...46L} {739, 46}

\bibitem[\protect\citeauthoryear{{{\L}okas}, {Athanassoula}, {Debattista}, {Valluri}, {Pino}, {Semczuk}, {Gajda}  \& {Kowalczyk}}{{{\L}okas} et~al.}{2014}]{Lokas2014}
{{\L}okas} E.~L.,  {Athanassoula} E.,  {Debattista} V.~P.,  {Valluri} M.,  {Pino} A.~d.,  {Semczuk} M.,  {Gajda} G.,   {Kowalczyk} K.,  2014, \mn@doi [\mnras] {10.1093/mnras/stu1846}, \href {https://ui.adsabs.harvard.edu/abs/2014MNRAS.445.1339L} {445, 1339}

\bibitem[\protect\citeauthoryear{{{\L}okas}, {Semczuk}, {Gajda}  \& {D'Onghia}}{{{\L}okas} et~al.}{2015}]{Lokas2015}
{{\L}okas} E.~L.,  {Semczuk} M.,  {Gajda} G.,   {D'Onghia} E.,  2015, \mn@doi [\apj] {10.1088/0004-637X/810/2/100}, \href {https://ui.adsabs.harvard.edu/abs/2015ApJ...810..100L} {810, 100}

\bibitem[\protect\citeauthoryear{Marinacci et~al.,}{Marinacci et~al.}{2018}]{Marinacci2018}
Marinacci F.,  et~al., 2018, \mn@doi [\mnras] {10.1093/mnras/sty2206}, 480, 5113

\bibitem[\protect\citeauthoryear{Martínez-García, del Pino, Aparicio, van der Marel  \& Watkins}{Martínez-García et~al.}{2021}]{MartinezGarcia2021}
Martínez-García A.~M.,  del Pino A.,  Aparicio A.,  van der Marel R.~P.,   Watkins L.~L.,  2021, \mn@doi [\mnras] {10.1093/mnras/stab1568}, 505, 5884

\bibitem[\protect\citeauthoryear{Martínez-García, del Pino  \& Aparicio}{Martínez-García et~al.}{2022}]{MartinezGarcia2022}
Martínez-García A.~M.,  del Pino A.,   Aparicio A.,  2022, \mn@doi [\mnras] {10.1093/mnras/stac3305}, 518, 3083

\bibitem[\protect\citeauthoryear{Mateo}{Mateo}{1998}]{Mateo1998}
Mateo M.,  1998, \mn@doi [\araa] {10.1146/annurev.astro.36.1.435}, 36, 435

\bibitem[\protect\citeauthoryear{Mayer}{Mayer}{2010}]{Mayer2010}
Mayer L.,  2010, \mn@doi [Advances in Astronomy] {10.1155/2010/278434}, 2010, 278434

\bibitem[\protect\citeauthoryear{McConnachie}{McConnachie}{2012}]{McConnachie2012}
McConnachie A.~W.,  2012, \mn@doi [\aj] {10.1088/0004-6256/144/1/4}, 144

\bibitem[\protect\citeauthoryear{McConnachie \& Venn}{McConnachie \& Venn}{2020}]{McConnachie2020a}
McConnachie A.~W.,  Venn K.~A.,  2020, \aj, 160, 124

\bibitem[\protect\citeauthoryear{{Mu{\~n}oz} et~al.,}{{Mu{\~n}oz} et~al.}{2006}]{Munoz2006}
{Mu{\~n}oz} R.~R.,  et~al., 2006, \mn@doi [\apj] {10.1086/505620}, \href {https://ui.adsabs.harvard.edu/abs/2006ApJ...649..201M} {649, 201}

\bibitem[\protect\citeauthoryear{Mu{\~{n}}oz et~al.,}{Mu{\~{n}}oz et~al.}{2005}]{Munoz2005}
Mu{\~{n}}oz R.~R.,  et~al., 2005, \mn@doi [\apj] {10.1086/497396}, 631, L137

\bibitem[\protect\citeauthoryear{Naiman et~al.,}{Naiman et~al.}{2018}]{Naiman2018}
Naiman J.~P.,  et~al., 2018, \mn@doi [\mnras] {10.1093/mnras/sty618}, 477, 1206

\bibitem[\protect\citeauthoryear{Navarro, Frenk  \& White}{Navarro et~al.}{1995}]{NavarroFrenkWhite1995}
Navarro J.~F.,  Frenk C.~S.,   White S. D.~M.,  1995, \mn@doi [\mnras] {10.1093/mnras/275.3.720}, 275, 720

\bibitem[\protect\citeauthoryear{Nelson et~al.,}{Nelson et~al.}{2018}]{Nelson2018}
Nelson D.,  et~al., 2018, \mn@doi [\mnras] {10.1093/mnras/stx3040}, 475, 624

\bibitem[\protect\citeauthoryear{Nelson et~al.,}{Nelson et~al.}{2019a}]{Nelson2019a}
Nelson D.,  et~al., 2019a, \mn@doi [Computational Astrophysics and Cosmology] {10.1186/s40668-019-0028-x}, 6, 2

\bibitem[\protect\citeauthoryear{Nelson et~al.,}{Nelson et~al.}{2019b}]{Nelson2019b}
Nelson D.,  et~al., 2019b, \mn@doi [\mnras] {10.1093/mnras/stz2306}, 490, 3234

\bibitem[\protect\citeauthoryear{Pace et~al.,}{Pace et~al.}{2020}]{Pace2020}
Pace A.~B.,  et~al., 2020, \mn@doi [\mnras] {10.1093/mnras/staa1419}, 495, 3022

\bibitem[\protect\citeauthoryear{Patel, Besla  \& Sohn}{Patel et~al.}{2017}]{Patel2017a}
Patel E.,  Besla G.,   Sohn S.~T.,  2017, \mn@doi [\mnras] {10.1093/mnras/stw2616}, 464, 3825

\bibitem[\protect\citeauthoryear{Patel, Besla, Mandel  \& Sohn}{Patel et~al.}{2018}]{Patel2018}
Patel E.,  Besla G.,  Mandel K.,   Sohn S.~T.,  2018, \mn@doi [\apj] {10.3847/1538-4357/aab78f}, 857, 78

\bibitem[\protect\citeauthoryear{Pillepich et~al.,}{Pillepich et~al.}{2018a}]{Pillepich2018a}
Pillepich A.,  et~al., 2018a, \mn@doi [\mnras] {10.1093/mnras/stx2656}, 473, 4077

\bibitem[\protect\citeauthoryear{Pillepich et~al.,}{Pillepich et~al.}{2018b}]{Pillepich2018b}
Pillepich A.,  et~al., 2018b, \mn@doi [\mnras] {10.1093/mnras/stx3112}, 475, 648

\bibitem[\protect\citeauthoryear{Pillepich et~al.,}{Pillepich et~al.}{2019}]{Pillepich2019}
Pillepich A.,  et~al., 2019, \mn@doi [\mnras] {10.1093/mnras/stz2338}, 490, 3196

\bibitem[\protect\citeauthoryear{{Planck Collaboration} et~al.,}{{Planck Collaboration} et~al.}{2016}]{PlanckCollaboration2016}
{Planck Collaboration} et~al., 2016, \mn@doi [A\&A] {10.1051/0004-6361/201525830}, 594, A13

\bibitem[\protect\citeauthoryear{Putman, Zheng, Price-Whelan, Grcevich, Johnson, Tollerud  \& Peek}{Putman et~al.}{2021}]{Putman2021}
Putman M.~E.,  Zheng Y.,  Price-Whelan A.~M.,  Grcevich J.,  Johnson A.~C.,  Tollerud E.,   Peek J. E.~G.,  2021, \mn@doi [\apj] {10.3847/1538-4357/abe391}, 913, 53

\bibitem[\protect\citeauthoryear{Rodriguez-Gomez et~al.,}{Rodriguez-Gomez et~al.}{2015}]{RodriguezGomez2015}
Rodriguez-Gomez V.,  et~al., 2015, \mn@doi [\mnras] {10.1093/mnras/stv264}, 449, 49

\bibitem[\protect\citeauthoryear{{Sales} et~al.,}{{Sales} et~al.}{2015}]{Sales2015}
{Sales} L.~V.,  et~al., 2015, \mn@doi [\mnras] {10.1093/mnrasl/slu173}, \href {https://ui.adsabs.harvard.edu/abs/2015MNRAS.447L...6S} {447, L6}

\bibitem[\protect\citeauthoryear{Springel}{Springel}{2010}]{Springel2010}
Springel V.,  2010, \mn@doi [\mnras] {10.1111/j.1365-2966.2009.15715.x}, 401, 791

\bibitem[\protect\citeauthoryear{Springel, White, Tormen  \& Kauffmann}{Springel et~al.}{2001}]{Springel2001}
Springel V.,  White S. D.~M.,  Tormen G.,   Kauffmann G.,  2001, \mn@doi [\mnras] {10.1046/j.1365-8711.2001.04912.x}, 328, 726

\bibitem[\protect\citeauthoryear{Springel et~al.,}{Springel et~al.}{2018}]{Springel2018}
Springel V.,  et~al., 2018, \mn@doi [\mnras] {10.1093/mnras/stx3304}, 475, 676

\bibitem[\protect\citeauthoryear{\VAN{van der Marel}{van der }{van der }{Marel} \& Cioni}{\VAN{van der Marel}{van der }{van der }{Marel} \& Cioni}{2001}]{VanderMarel2001}
\VAN{van der Marel}{van der }{van der }{Marel} R.~P.,  Cioni M.-R.~L.,  2001, \mn@doi [\aj] {10.1086/323099}, 122, 1807

\bibitem[\protect\citeauthoryear{\VAN{van der Marel}{van der }{van der }{Marel}, Alves, Hardy  \& Suntzeff}{\VAN{van der Marel}{van der }{van der }{Marel} et~al.}{2002}]{VanderMarel2002}
\VAN{van der Marel}{van der }{van der }{Marel} R.~P.,  Alves D.~R.,  Hardy E.,   Suntzeff N.~B.,  2002, \mn@doi [\aj] {10.1086/343775}, 124, 2639

\bibitem[\protect\citeauthoryear{Walker, Mateo  \& Olszewski}{Walker et~al.}{2009}]{Walker2009}
Walker M.~G.,  Mateo M.,   Olszewski E.~W.,  2009, \mn@doi [\aj] {10.1088/0004-6256/137/2/3100}, 137, 3100

\bibitem[\protect\citeauthoryear{Weinberger et~al.,}{Weinberger et~al.}{2017}]{Weinberger2017}
Weinberger R.,  et~al., 2017, \mn@doi [\mnras] {10.1093/mnras/stw2944}, 465, 3291

\bibitem[\protect\citeauthoryear{{Weisz}, {Dolphin}, {Skillman}, {Holtzman}, {Gilbert}, {Dalcanton}  \& {Williams}}{{Weisz} et~al.}{2014}]{Weisz2014}
{Weisz} D.~R.,  {Dolphin} A.~E.,  {Skillman} E.~D.,  {Holtzman} J.,  {Gilbert} K.~M.,  {Dalcanton} J.~J.,   {Williams} B.~F.,  2014, \mn@doi [\apj] {10.1088/0004-637X/789/2/147}, \href {https://ui.adsabs.harvard.edu/abs/2014ApJ...789..147W} {789, 147}

\bibitem[\protect\citeauthoryear{Wheeler et~al.,}{Wheeler et~al.}{2017}]{Wheeler2017}
Wheeler C.,  et~al., 2017, \mn@doi [\mnras] {10.1093/mnras/stw2583}, 465, 2420

\bibitem[\protect\citeauthoryear{White \& Rees}{White \& Rees}{1978}]{WhiteRees1978}
White S. D.~M.,  Rees M.~J.,  1978, \mn@doi [\mnras] {10.1093/mnras/183.3.341}, 183, 341

\bibitem[\protect\citeauthoryear{Wilkinson, Kleyna, Evans, Gilmore, Irwin  \& Grebel}{Wilkinson et~al.}{2004}]{Wilkinson2004}
Wilkinson M.~I.,  Kleyna J.~T.,  Evans N.~W.,  Gilmore G.~F.,  Irwin M.~J.,   Grebel E.~K.,  2004, \mn@doi [\apj] {10.1086/423619}, 611, L21

\bibitem[\protect\citeauthoryear{Zhu, van~de Ven, Watkins  \& Posti}{Zhu et~al.}{2016}]{Zhu2016}
Zhu L.,  van~de Ven G.,  Watkins L.~L.,   Posti L.,  2016, \mn@doi [\mnras] {10.1093/mnras/stw2081}, 463, 1117

\makeatother
\end{thebibliography}

%\begin{thebibliography}{99}
%\end{thebibliography}

%%%%%%%%%%%%%%%%%%%%%%%%%%%%%%%%%%%%%%%%%%%%%%%%%%

%%%%%%%%%%%%%%%%% APPENDICES %%%%%%%%%%%%%%%%%%%%%

%\appendix


%%%%%%%%%%%%%%%%%%%%%%%%%%%%%%%%%%%%%%%%%%%%%%%%%%

% Don't change these lines
\bsp	% typesetting comment
\label{lastpage}
\end{document}